\documentstyle[12pt,epsf]{article} %XXX: what is esbbsm???
\def\doref#1#2#3{#1, #2 (#3)}
\def\semi{;\\}
\def\etal{{\it et al.}}
\def\ibid{{\it ibid}}

\def\npb#1#2#3{Nucl. Phys. B \doref{#1}{#2}{#3}}
\def\plb#1#2#3{Phys. Lett. B \doref{#1}{#2}{#3}}
\def\prl#1#2#3{Phys. Rev. Lett. \doref{#1}{#2}{#3}}
\def\pr#1#2#3{Phys. Rev. \doref{#1}{#2}{#3}}

\def\prd#1#2#3{Phys. Rev. D \doref{#1}{#2}{#3}}
\def\prp#1#2#3{Phys. Rep. \doref{#1}{#2}{#3}}

\def\zpc#1#2#3{Z. Phys. C \doref{#1}{#2}{#3}}
\def\ptp#1#2#3{Prog. Theor. Phys. \doref{#1}{#2}{#3}}
\def\ap#1#2#3{Ann. Phys. (NY) \doref{#1}{#2}{#3}}
\def\arnp#1#2#3{Ann. Rev. Nucl. Part. Sci \doref{#1}{#2}{#3}}
\def\mpla#1#2#3{Mod. Phys. Lett. A \doref{#1}{#2}{#3}}

\def\rmp#1#2#3{Rev. Mod. Phys. \doref{#1}{#2}{#3}}
\def\app#1#2#3{Acta. Phys.  Polon. \doref{#1}{#2}{#3}}
\def\pha#1#2#3{Physica \doref{#1}{#2}{#3}}
\def\nim#1#2#3{Nucl. Inst. and Meth. \doref{#1}{#2}{#3}}
\def\cpc#1#2#3{Comp. Phys. Commun. \doref{#1}{#2}{#3}}
\def\rpp#1#2#3{Rept. Prog. Phys. \doref{#1}{#2}{#3}}
\def\hepph#1{hep-ph/#1}

\def\gv{\tilde{g}}

\def\Tr{{\rm Tr}}

\def\gs{g''}
\def\rs{\sqrt{s}}

\newcommand{\be}{\begin{equation}}
\newcommand{\ee}{\end{equation}}
\newcommand{\bea}{\begin{eqnarray}}
\newcommand{\eea}{\end{eqnarray}}

\def\ie{{\it i.e.}}
\def\9{\phantom 0}      %%% for lining up numbers in columns
\renewcommand\linebreak{\unskip\break} %% breaks line & still justifies
\def\Tr{{\rm Tr}}

\def\gs{g''}
\def\rs{\sqrt{s}}

\def\half{{1 \over 2}}

\def\punct[#1]{~~#1}

\def\journal{\topmargin .3in    \oddsidemargin .5in
        \headheight 0pt \headsep 0pt
        \textwidth 5.625in % 1.2 preprint size  %6.5in
\textheight 8.25in % 1.2 preprint size 9in
        \marginparwidth 1.5in
        \parindent 2em
        \parskip .5ex plus .1ex         \jot = 1.5ex}
\journal

\newskip\humongous \humongous=0pt plus 1000pt minus 1000pt

\newif\ifdtup

\def\ww{W^+W^-}
\def\wwl{W_{\rm L}^+W_{\rm L}^-}
\def\Tr{{\rm Tr~}}
\def\L{{\cal L}}
\def\V{{\cal V}}
\def\A{{\cal A}}
\def\CM{{\cal M}}

\def\ali{a^I_\ell}
\def\tli{t^I_\ell}
\def\frac#1#2{{#1 \over #2}}

\newcommand{\mZ}{\mbox{$m_Z$}}

\newcommand{\SB}{\mbox{S/$\sqrt{B}$}}

\newcommand{\WL}{\mbox{$W_L$}}
\newcommand{\ZL}{\mbox{$Z_L$}}

\def\pT{$p_T$}
\def\mH{$m_H$}
\def\abseta{$|\eta|$}
\def\dR{$\delta R$}
\def\detaphi{$\delta \eta \delta \phi$}

\def\undertext#1{\vtop{\hbox{#1}\kern 1pt \hrule}}

\def\wlwl{$W^+_L W^+_L\ $}
\def\wtwt{$W^+_T W^+_T\ $}

\def\ttbar{$t {\overline t}$}
\def\HWW{$H \to W^+W^-$}
\def\HZZ{$H \to ZZ$}
\def\HWWlnj{$H \to W^+W^- \to \ell \nu$~jets}
\def\HZZllj{$H \to ZZ \to \ell^+ \ell^-$~jets}
\def\HZZl{$H \to ZZ \to \ell^+ \ell^- \ell^+ \ell^-$}
\def\agt{\raise-.5ex\vbox{\hbox{$\; >\;$}\vskip-2.9ex\hbox{$\; \sim\;$}}}
\begin{document}
\begin{flushright}
        UCD-95-32\\
        ISU-HET-95-6\\
        October 1995
\end{flushright}
\newlength{\captsize} \let\captsize=\small % use \let\normalsize=\captsize
\newlength{\captwidth}                     % just before \caption{ ...
%%%%%%%%%%%%%%%%%%%%%%%%%%%%%%%%%%%%%%%%%%%%%%%%%%%%%%%%%%%%%%
%
\begin{center}
\vskip6pt
\bf
STRONGLY-INTERACTING ELECTROWEAK SECTOR\\
--MODEL INDEPENDENT APPROACHES--\\
\rm
\vskip1pc
M.~Golden$^{(a)}$, T.~Han$^{(b)}$ and G.~Valencia$^{(c)}$\\
$^{(a)}${\em Lyman Laboratory of Physics,}
{\em Harvard University, Cambridge, MA 02138}\\
E-mail:  golden@neographic.com \\
$^{(b)}${\em Department of  Physics,
University of California, Davis, CA 95616}\\
E-mail:  than@ucdhep.ucdavis.edu \\
$^{(c)}${\em Department of Physics,
Iowa State University, Ames, IA 50011}\\
E-mail:  valencia@iastate.edu \\
\end{center}
\vskip2pc

\tableofcontents

\section{Introduction}

One of the open questions in high energy physics is the mechanism of
electroweak symmetry breaking.  The physics that breaks electroweak
symmetry is responsible for giving the $W$ and $Z$ gauge bosons
their masses, removing the symmetry that connects them to the
massless photon.  Since a massive spin-one particle has three
polarizations, rather than the two of a massless mode, the new
physics must supply degrees of freedom to be swallowed by the gauge
bosons.

In an approximate sense, these swallowed degrees of freedom are the
longitudinal polarizations $W_L$ and $Z_L$ of the gauge bosons.  If we
examine any amplitude for a subprocess involving longitudinal gauge
bosons, we find that as the energy gets large, the $W_L$ and $Z_L$
behave more and more like the original swallowed degrees of freedom
\cite{equivth,lqt,changail}.  Therefore, a good way to probe the
interactions of the symmetry breaking sector is by the interactions
of the longitudinal components of the gauge bosons.

Over the years many competing ideas for the electroweak symmetry
breaking sector have been suggested.  Some of these, such as
technicolor models \cite{techni,technirev}, include strong self-interactions.
In fact, strong self-interactions are a fairly generic feature of
the models with dynamical symmetry breaking.

Strongly self-coupled models are difficult to solve using analytic
means.  However, any proposed model of the electroweak symmetry
breaking sector must produce three Goldstone bosons, to be swallowed
by the $W$ and $Z$.  By dint of the fact that they are Goldstone
bosons, the longitudinal $W$ and $Z$ can be described by chiral
Lagrangian techniques \cite{ccwz,appel}, and certain low-energy
theorems \cite{loweth,CGG} can be shown to hold for their interactions.
These theorems hold independent of any particular structure of the
high energy theory.  The Goldstone bosons may be fundamental (as in
the strongly self-coupled standard model) or composite (as in
technicolor) - their interactions near threshold are determined
entirely from the symmetry structure of the theory.

In this chapter we explore the electroweak symmetry breaking sector
using the approach outlined above.  The goal is to see what
statements can be made about experiments that probe the symmetry
breaking sector without making reference to the detailed
interactions of the underlying theory.

As described below, the interactions of the longitudinal gauge
bosons grow with energy.  As we shall see, if no new physics enters
to cut off their growth, at energies of order 1 TeV or so the
longitudinal gauge bosons are so strongly self-coupled that the
scattering amplitudes would violate unitarity \cite{lqt}.  One
expects therefore that some sort of new physics comes in to cut off
the growth in the amplitudes.  Frequently this new physics takes the
form of new resonances with masses near the TeV scale.  For example,
in the standard model (SM), a neutral scalar Higgs boson is
present \cite{lighth}.

There is, however, nothing especially sacrosanct about the Higgs
scalar -- there are many possibilities for this new physics, and
which one is chosen will depend on the details of the underlying
dynamics.  Such a strongly-interacting symmetry breaking sector need
not be at all like the minimal standard model, and many alternatives
have been suggested. For example, one could have a richer
spectrum of resonances in the few TeV region, as is expected for
technicolor models \cite{technirev}.

If one has a specific model for a strongly-interacting electroweak
symmetry breaking sector, one can, at least in principle, predict
the resonances that should appear in high energy experiments.  That,
however, would take us beyond the scope of this chapter
and is the subject of another working group \cite{tcgroup}.
Here, we
concern ourselves with the phenomenology of a generic
strongly-interacting electroweak symmetry breaking sector at future
colliders.  However, we will adopt a phenomenological approach,
describing the experimental signals for different kinds of new
physics.  Thus we go a little bit, but not too much, beyond strict
``model independence''.

We will parameterize three simple scenarios: one in which there are
no resonances in the experimentally accessible region; one in which
the physics is dominated by a new particle with the quantum numbers
of the Higgs boson; and one in which the physics is dominated by a
new vector particle.

In very high energy processes, there are many competing mechanisms
to produce $W$ and $Z$ bosons.  Some of these are more sensitive than
others to the longitudinal components. A complete numerical study of
a strongly-interacting symmetry breaking sector
beyond the SM has not been done.
Rather, some simplifying assumptions have generally been used.  For
example, in this chapter we isolate the kinematic region at fairly
large energy, in which the interactions of longitudinal gauge bosons
dominate. If instead one concentrates exclusively on low energy
physics, one can choose to focus on the changes made by new physics
to the gauge boson self-couplings.  This simplified scenario is
studied in the chapter on Anomalous Gauge Boson Interactions \cite{anon}.

For the study of processes in which the longitudinal components of
the $W$ and $Z$ dominate the scattering amplitudes, it is sufficient
to use the Equivalence Theorem (ET) \cite{equivth,lqt,changail,eqth,hky}
to extract from the full amplitude
only those terms that are of ``enhanced electroweak strength''. This
means that one can calculate amplitudes replacing all $W$ and $Z$
gauge bosons by their corresponding would-be Goldstone bosons in
Landau gauge. By doing this one obtains amplitudes that are correct
up to terms of order ${\cal O}(M_W/E)$. Results for $WW$ scattering
in this approximation are only valid for $s \gg M^2_W$.
%In a full
%numerical study, this condition would be relaxed, and there would be
%no distinction between the processes studied here and those of the
%chapter on Anomalous Couplings \cite{lukef}.

One mechanism to produce vector boson pairs at future colliders is
through light fermion anti-fermion annihilation.  This is the case
of light $q \bar{q}$ annihilation in hadronic colliders
(the Drell-Yan mechanism), as well as
the case of $e^+ e^-$ annihilation in future $e^+ e^-$ colliders.
This
process yields vector boson pairs that are mostly transversely
polarized and will usually be a background to the processes
considered here.  The one important exception is the production of
longitudinally polarized vector bosons in a $J=1$
state \cite{changail,Chan,BESSssc,baggeretali,mcwk,pseudo,bc,ggvv,themany}.
This production mechanism is thus very
sensitive to new physics with a vector resonance like a techni-rho.

A second mechanism for producing longitudinal vector boson pairs in
hadronic colliders is gluon fusion \cite{gfusion}. In this case the
initial gluons turn into two vector bosons via an intermediate state
that couples to both gluons and electroweak gauge bosons.  Examples
are: the top quark, new heavy quarks, and new colored particles of a
technicolor model.
%The first case is discussed in the chapter on
%anomalous couplings of the top-quark; and the third case in the
%chapter on the implications of specific models.  We remind the
%reader that, i
In this case, only chargeless $V_L V_L$ pairs can be
produced, and thus this channel is particularly sensitive to new
physics with a scalar resonance like a heavy Higgs boson.

Finally, there is the vector-boson fusion process
\footnote{The gauge-boson fusion mechanism
was first discussed in the context of $e^+e^-$ scattering in
\cite{vvfusee,vvfuse}.  For $eN$ scattering, it was discussed in
\cite{vvfuseN,vvfuse}. Gauge-boson scattering in $pp$ collisions was first
discussed in \cite{vvfuspp,vvfuse}.},
say $V_L^{}V_L^{} \to V_L^{}V_L^{}$, which is especially
important in the case of a strongly-interacting electroweak
symmetry breaking sector.
The major advantage for studying the vector-boson fusion processes
is that they involve all possible spin and isospin channels
simultaneously, with scalar and vector resonances as well as
non-resonant channels.

In our phenomenological discussions below,
we will mainly concentrate on the first and the
last mechanisms, due to the consideration of signal identification
and background suppression.

%\subsection{Global Symmetries and Unitarity}
\subsection{Global Symmetries}

Let us begin by recalling that in the Standard Model, the $W_LW_L$
scattering amplitudes are unitarized by exchange of a spin-zero resonance,
the Higgs particle $H$.  The Higgs boson
is contained in a complex scalar doublet,
\begin{equation}
\Phi\ =\ (v + H) \exp(i w^a \sigma^a/v)\ ,
\label{threeone}
\end{equation}
where the $\sigma^a$ are the conventional Pauli matrices.
The four components of
$\Phi$ contain three would-be Goldstone bosons $w^a$
and the Higgs particle $H$.  In the Standard Model, the Higgs
potential,
\begin{equation}
\V\ =\ {\lambda \over 16}\,
\bigg[ {\rm Tr}\,\big(\Phi^\dagger\Phi - v^2\big)\bigg]^2,
\label{threetwo}
\end{equation}
is invariant under a global $SU(2)_L \times SU(2)_R$ symmetry,
\begin{equation}
\Phi\ \to\ L\,\Phi\,R^\dagger\ ,
\label{threethree}
\end{equation}
with $L,R \in SU(2)$.  The vacuum expectation value
\begin{equation}
\langle\Phi\rangle \ =\  v\ ,
\label{threefour}
\end{equation}
breaks the symmetry to the diagonal $SU(2)$.  In the perturbative limit,
it also gives mass to the Higgs boson,
\begin{equation}
m_H\ =\ \sqrt{2 \lambda}\, v\ ,
\label{threefive}
\end{equation}
where $v=246$ GeV.

In the Standard Model, the diagonal $SU(2)$ symmetry is broken only
by terms proportional to the hypercharge coupling $g'$ and the
up-down fermion mass splittings.  It is responsible for the
successful mass relation
\begin{equation}
M_W\ =\ M_Z\ \cos\theta\ ,
\label{threesix}
\end{equation}
where $\theta$ is the weak mixing angle; $M_W$ and $M_Z$ are the masses of
$W^\pm$ and $Z$, respectively.
The four components of $\Phi$ split into a triplet $w^a$ and a singlet
$H$ under the unbroken diagonal $SU(2)$ symmetry.
In analogy to the chiral symmetry of
QCD, we call the unbroken $SU(2)$ ``isospin''
(or ``custodial symmetry'') \cite{rhoparameter}.

Beyond the Standard Model, the electroweak interactions can be broken
by an arbitrary symmetry breaking sector. In all cases, however,
we want the electroweak gauge group $SU(2)_L \times U(1)_Y$ to be
spontaneously broken to $U(1)_Q$. The minimal global symmetry
consistent with this pattern of symmetry breaking is thus a
global $SU(2) \times U(1)$ that is broken spontaneously to $U(1)$.
The global symmetry group can also be larger than this one.
For example it could be $SU(2) \times SU(2)$
that breaks down to $SU(2)$, as in the minimal Standard Model.
As discussed above, in this case there is an isospin (or custodial)
$SU(2)$ symmetry. We will restrict our study to interactions that
respect this symmetry.

Of course, the global symmetry group may be larger, as is the case in
some technicolor models \cite{onef}. When this happens, however, there are
pseudo-Goldstone bosons remaining in the physical spectrum. We will
not consider the possibilities for direct observation of the
(model-dependent) pseudo-Goldstone bosons.

It is possible to study some aspects of electroweak symmetry breaking
in a model independent way using the language of effective (``chiral'')
Lagrangians. There is some information that will
be common to all theories of electroweak symmetry breaking and determines
the low energy behavior of scattering amplitudes.

Introducing the would-be Goldstone boson fields, $w^+$, $w^-$,
and $z$ through the matrix
\begin{equation}
\Sigma = \exp{(i \vec{\sigma}\cdot \vec{w}/v)}\ ,
\label{sigdef}
\end{equation}
we can construct an effective Lagrangian that contains this very general
information:
\begin{equation}
{\cal L}= {v^2 \over 4} {\rm Tr}\partial_{\mu}\Sigma \partial^{\mu}
\Sigma^{\dagger}
\label{lola}
\end{equation}
This is the most general effective Lagrangian one can write with only
two derivatives to describe the interactions of the Goldstone-bosons
associated with the spontaneously broken global symmetry $SU(2)\times SU(2)
\to SU(2)$. The gauge interactions of the standard model
are introduced by requiring the Lagrangian in Eq.~(\ref{lola}) to be
gauge invariant under $SU(2)_L \times U(1)_Y$. This is accomplished by
replacing the derivative with a covariant derivative:
\begin{equation}
\partial_{\mu}\Sigma \to {\cal D}_{\mu} \Sigma =
\partial_{\mu}\Sigma -i{g\over 2}W^{\alpha}_{\mu}\sigma^{\alpha}\Sigma
+ i {g^{\prime} \over 2}B_{\mu}\Sigma \sigma_3^{}
\label{cov}
\end{equation}
This construction is thus an $SU(2)_L \times U(1)_Y$ gauge invariant mass
term for the $W$ and $Z$ satisfying the relation Eq.~(\ref{threesix}).

\subsection{$V_L V_L$ Scattering Amplitudes}

At high energies, the scattering of longitudinally polarized vector
bosons ($V_L$)
can be approximated by the scattering of the would-be
Goldstone bosons $w^a$
\cite{equivth,lqt,changail,CGG,eqth,hky,MGIowa}.  For the
Standard Model, this is a calculational simplification, but for
other models it is a powerful conceptual aid as well.  For example,
if one thinks of the would-be Goldstone fields in analogy with the
pions of QCD, one expects the $W_LW_L$ scattering amplitudes to be
unitarized by a spin-one, isospin-one vector resonance, like the
techni-rho.  Alternatively, if one thinks of the Goldstone fields in
terms of the linear sigma model, one expects the scattering
amplitudes to be unitarized by a spin-zero, isospin-zero scalar
field like the Higgs boson.

In this chapter, we are interested in the strongly-interacting
longitudinal $W$'s in the TeV region.  We will ignore the gauge
couplings and the up-down fermion mass splittings.  Therefore, the
SU(2) ``isospin'' is conserved.  The $W_LW_L$ scattering amplitudes
can then be written in terms of isospin amplitudes, exactly as in
low energy hadron physics. We assign isospin indices as follows,
\begin{equation}
W^a_L\ W^b_L\ \to\  W^c_L\ W^d_L\ ,
\label{threeseven}
\end{equation}
where $W_L$ denotes either $W_L^\pm$ or
$Z_L$, where $W_L^\pm=(1/\sqrt{2})(W_L^1 \mp iW_L^2)$ and
$Z_L=W_L^3$.  The scattering amplitude is given by
\begin{equation}
\CM(W^a_LW^b_L\to W^c_LW^d_L)\ =\ A(s,t,u)\delta^{ab}
\delta^{cd}\ +\ A(t,s,u)\delta^{ac}\delta^{bd}\ +
\ A(u,t,s)\delta^{ad}\delta^{bc}\ ,
\label{threeeight}
\end{equation}
where $a,b,c,d=1,2,3$, and $s$, $t$, and $u$ are the usual
Mandelstam variables.  All the physics of $W_LW_L$ scattering is
contained in the amplitude function $A$.

Given the amplitude functions, the physical amplitudes for
boson-boson scattering are given as follows,
\begin{eqnarray}
\CM(W^+_LW^-_L\to Z_LZ_L)\ & = &\ A(s,t,u) \nonumber \\
\CM(Z_LZ_L\to W^+_LW^-_L)\ & = &\ A(s,t,u) \nonumber \\
\CM(W^+_LW^-_L\to W^+_LW^-_L)\ & = &\ A(s,t,u)\ +\ A(t,s,u) \nonumber \\
\CM(Z_LZ_L\to Z_LZ_L)\ & = &\ A(s,t,u)\
+\ A(t,s,u)\ +\ A(u,t,s) \nonumber \\
\CM(W^\pm_LZ_L\to W^\pm_LZ_L)\ & = &\ A(t,s,u) \nonumber \\
\CM(W^\pm_LW^\pm_L\to W^\pm_LW^\pm_L)\ & = &\ A(t,s,u)\ +
\ A(u,t,s)\ .
\label{threenine}
\end{eqnarray}

\section{Strongly-interacting Electroweak Symmetry Breaking} %Sector}

For the Standard Model, the amplitude functions are easy to
work out.  They can be expressed by

\begin{equation}
A(s,t,u)\ =\ {-m_H^2\over v^2}\left(1 + {m_H^2\over s-m_H^2 +
i m_H \Gamma_H \theta (s)}\right)\ ,
\label{threetwelve}
\end{equation}
where $m_H$ and $\Gamma_H = 3 m^3_H/32 \pi v^2$ are the
mass and width of the Higgs boson; $\theta (s)$ is the step-function
which takes the value one for $s > 0$ and zero otherwise.\footnote{
Here and henceforth, we include a constant Breit-Wigner width
for the Higgs particle in the $s$-channel,
there are several subtleties associated
with this \cite{rwwll,ny,seym}.
We do not include the width in the non-resonant channels.}

If we look at the Standard Model amplitude for $WW$ scattering in the
$I=0$ channel we find:
\begin{equation}
{\cal M} = {1 \over v^2}\bigg( 3s+t+u - {3 s^2 \over s-M^2_H}
-{t^2 \over t-M^2_H} -{u^2 \over u-M^2_H} \biggr)
\label{lowsm}
\end{equation}
and if we project out the $J=0$ partial wave, we obtain for $s \gg M^2_H$:
\begin{equation}
a^0_0 \to {5 M^2_H \over 32 \pi v^2}
\end{equation}
a result that is proportional to $M^2_H$.  However, a minimal
consequence of partial wave unitarity is that
\begin{equation}
|{\rm Re}a^0_0| \leq {1 \over 2}
\label{unidef}
\end{equation}
So we can see that for a sufficiently large Higgs boson mass (about
800 $GeV$), this amplitude will ``violate unitarity''. Of course
no properly calculated amplitude will ever violate unitarity, so
what this means is that
the amplitude is becoming sufficiently large that we cannot trust our
perturbative calculation. This has been interpreted to mean that the
minimal standard model becomes strongly-interacting for a sufficiently
heavy Higgs boson.
This is the simplest example of a strongly-interacting
electroweak symmetry breaking sector.

Similarly, the lowest order expansion of Eq.~(\ref{lowsm}) (the terms
quadratic in energy) is reproduced by the effective Lagrangian of
Eq.~(\ref{lola}).  Since it follows just from the pattern of
symmetry breaking, this low energy behavior is universal.

In what follows, we discuss effective theories,
based on nonrenormalizable effective Lagrangians
for the strong $WW$ sector.  These models must be understood in the context
of an energy expansion. Generally, such an expansion does
not provide a unitary description for all energies. This is simply because
the effective Lagrangian does not make explicit the new physics that must
appear at some scale $\Lambda$, well above the $WW$ mass region where it
is to be employed. For numerical estimates it is often necessary to
cut off the bad high energy behavior of the scattering amplitudes
outside the region of validity of the effective theory. This introduces
some undesirable but unavoidable systematic error. Several such
unitarization schemes have been applied in the literature, for example:
the $K$-matrix unitarization, Pad\`{e} approximants,
and $N/D$ methods \cite{unitar}. In our discussions, we will
follow some simple treatments, either the $K$-matrix unitarization
or some cutoff scheme (see following sections).

\subsection{Nonresonant Models}

Effective field theories can describe nonresonant models in which
the $W_LW_L$ scattering occurs below the threshold for resonance
production \cite{appel}.  The effective Lagrangian description allows
one to construct gauge invariant scattering amplitudes that are
consistent with the global symmetries, crossing symmetry, and
perturbative unitarity order by order in an energy
expansion \cite{nrm}.

The lowest order effective Lagrangian, Eq.~(\ref{lola}), gives the
universal, leading order, behavior of the scattering amplitudes.  In
order to calculate amplitudes to next to leading order in powers of
the external momenta, $p^4$, one uses Eq.~(\ref{lola}) at tree and
one-loop levels, and the most general effective Lagrangian
consistent with the desired symmetries that contains four
derivatives. The infinities that appear when using Eq.~(\ref{lola})
at one loop, are absorbed by defining renormalized parameters in the
next to leading order effective Lagrangian.

At next to leading order one finds corrections to the lowest order
behavior of amplitudes.  Unlike the lowest order computation, these
actually depend on the underlying dynamics. This description of
scattering amplitudes has been seen to work reasonably well for the
case of $\pi \pi$ scattering up to energies of about 500 MeV.  In
QCD, the scale of chiral symmetry breaking (that is, the scale by
which the higher dimension operators are suppressed) is about
1 GeV; if the electroweak symmetry breaking sector is analogous
then we may naively scale the scattering amplitudes by a factor of
$v/f_\pi$ and expect the effective Lagrangian description of $ww$
scattering to be ``reasonable'' below about 1.5 TeV.

This type of model allows the study of $WW$ scattering in a ``low
energy region'', below threshold for production of any new
resonance. In this way we can assess the capability of a new
collider to study electroweak symmetry breaking when it cannot
directly produce new heavy resonances.

Since we limit our discussion to the scattering of longitudinal
vector bosons at high energy, it is sufficient to consider the
effective Lagrangian for the Goldstone fields.  We note that the
field $\Sigma$ defined in Eq. (\ref{sigdef}) transforms as
\begin{equation}
\Sigma \to L \Sigma R^\dagger
\end{equation}
under $SU(2)_L \times SU(2)_R$.  The most general $SU(2)_L \times
SU(2)_R$ invariant Lagrangian for the Goldstone fields
containing four or fewer derivatives is \cite{appel}
\begin{eqnarray}
\L_{\rm Goldstone} \ & = &  {v^2 \over 4}\,
\Tr \partial_\mu \Sigma^\dagger
\partial^\mu \Sigma \nonumber \\
 &\ +&\ L_1\,\bigg({v \over \Lambda}\bigg)^2\,
 \Tr(\partial_\mu\Sigma^\dagger \partial^\mu \Sigma)
 \ \Tr(\partial_\nu\Sigma^\dagger \partial^\nu \Sigma) \nonumber \\
& \ +&\ L_2\,\bigg({v \over \Lambda}\bigg)^2\,\Tr(\partial_\mu\Sigma^\dagger
\partial_\nu \Sigma)\ \Tr(\partial^\mu\Sigma^\dagger \partial^\nu \Sigma)\ ,
\label{goldlag}
\end{eqnarray}
where $\Lambda \leq 4 \pi v$ denotes the scale of the new physics
\cite{NDA}.

When the effective Lagrangian is gauged, there are additional terms
that can be added.  The coefficients of these extra terms induce
``oblique'' corrections \cite{oblique} to the gauge boson
propagators, and anomalous three- and four-gauge boson
couplings \cite{anon}.

With the effective Lagrangian, Eqs.~(\ref{lola}) and (\ref{goldlag}),
to order $p^4$, the scattering amplitudes are given by \cite{appel}
\begin{eqnarray}
A(s,t,u) \ & = &  {s \over v^2}\ +
\ {1 \over 4 \pi^2 v^4}\ \biggl( 2\, L_1(\mu)\,s^2
\ +\ L_2(\mu)\, (t^2+u^2)\ \biggr)  \nonumber \\
&\ +&\ {1 \over 16 \pi^2 v^4}\ \biggl[-{t \over 6}\,(s+2t){\rm log}
\biggl(-{t \over \mu^2}\biggr)\ -\ {u \over 6}\,(s+2u){\rm log}
\biggl(-{u \over \mu^2}\biggr)\  \nonumber \\
&\ -&\ {s^2 \over 2}\, {\rm log}\biggl(-
{s \over \mu^2} \biggr)\biggr]\ ,
\label{goldamps}
\end{eqnarray}
where we have taken $\Lambda = 4 \pi v \sim 3.1$ TeV and the
$L_i(\mu)$ are the renormalized coefficients in the effective
Lagrangian.  (${\rm log}(-s) = {\rm log}(s)-i \pi$, for $s > 0$.)
To this order, there are two types of contributions.  The first is a
direct coupling that follows from the tree-level Lagrangian.  The
second is a one-loop correction that must be included at order
$p^4$.  The loop contribution renormalizes the parameters $L_1$ and
$L_2$, and gives finite logarithmic corrections that cannot be
absorbed into a redefinition of the couplings. The parameters
$L_{1,2}$ contain the information about the physics that breaks
electroweak symmetry.

One difficulty with this low-energy effective Lagrangian approach is
that the scattering amplitudes violate unitarity for $WW$ invariant masses
between 1 and 2 TeV.  This indicates that the low energy description is
breaking down because new physics is near. Since the new colliders do
not provide monochromatic beams of $W$ bosons, but rather a continuous
spectrum, in any practical calculation we are forced to deal with
scattering amplitudes at energies outside the region of validity of the
low-energy description. The standard treatment of this problem is the
unitarization of the scattering amplitudes. Unfortunately this procedure
is not unique and introduces systematic errors in the studies.

Some unitarization prescriptions that have been used in the
literature include:

\begin{itemize}

\item{1)}  Take $L_1(\mu) = L_2(\mu) = 0$ and ignore the loop-induced
logarithmic corrections to the scattering amplitudes.  The resulting
amplitudes are universal in the sense that they depend only on $v$.
They reproduce the low-energy theorems of pion dynamics.  Unitarize
these amplitudes by saturating the partial waves when they reach the
bound $|\ali| \leq 1$.  This is the original model considered by
Chanowitz and Gaillard \cite{changail}, so we call it {\it LET CG}.

\item{2)}  Another approach is to take $L_1(\mu) = L_2(\mu) = 0$,
ignore the loop-induced logarithmic corrections as before, and
unitarize the scattering amplitudes using a ``$K$-matrix.''  That
is, replace the partial wave amplitudes $\ali$ by $\tli$, where
\begin{equation}
\tli\ =\ {\ali \over 1-i\ali }\ .
\label{tamps}
\end{equation}
Note that neither this method nor method 1) preserves crossing
symmetry.  We call this model {\it LET K}.

\item{3)}  A third possibility includes the full ${\cal O}(p^4)$
amplitude presented above.  By varying the parameters $L_1(\mu)$ and
$L_2(\mu)$, one can sweep over all possible nonresonant physics.  In
particular, one can search for a region where equation
(\ref{unidef}) is not violated before 2 TeV.  Scanning the
($L_1(\mu), L_2(\mu)$) parameter space, one finds that the values
\cite{bdvs}
\begin{eqnarray}
L_1(\mu)\ & = &\ -0.26 \nonumber \\
L_2(\mu)\ & = &\ +0.23 \ ,
\label{loneltwo}
\end{eqnarray}
measured at the renormalization scale $\mu=1.5$ TeV, maximize the
scale at which (\ref{unidef})\ breaks down.  Beyond 2 TeV, the partial
waves are no longer unitary.  In order to compare with the total
event rates in the other models, above 2 TeV we unitarize the
scattering amplitudes using the $K$-matrix prescription, so we call
this model {\it DELAY K}.  Note that only the real part of
$a^I_\ell$ in Eq.~(\ref{goldamps}) is used to obtain the unitarized
partial wave amplitude $t^I_\ell$ above 2 TeV.

\end{itemize}

\subsection{Spin-zero, Isospin-zero Resonances}

We now consider the case where the symmetry breaking sector has a
scalar resonance as its dominant feature at energies up to a few
TeV.  This corresponds, for example, to a standard model Higgs boson
in the non-perturbative regime, or to a technicolor-like theory
whose lowest resonance is a techni-sigma.  As before, we may
construct an effective Lagrangian, which will be consistent with the
chiral symmetry $SU(2)_L \times SU(2)_R$, spontaneously broken to
the diagonal $SU(2)$.

The basic fields are $\Sigma$ and a scalar $S$.  The new field $S$
transforms as a singlet under $SU(2)_L \times SU(2)_R$,
\begin{equation}
S \to S\ .
\end{equation}

This is all we need to construct the effective Lagrangian.  To the
order of interest, it is given by
\begin{eqnarray}
\L_{\rm Scalar}\ & = &\ \ {v^2 \over 4}\,
\Tr \partial^\mu \Sigma^\dagger
\partial_\mu \Sigma \nonumber \\
&& \ +\ {1\over2}\,\partial^\mu S \partial_\mu S \ -
\ {1\over2}\,M^2_S\,S^2 \nonumber \\
&& \ +\ {1\over2}\,gv\,S\,\Tr \partial^\mu \Sigma^\dagger \partial_\mu
\Sigma\ +\ \ldots \ ,
\label{lscalar}
\end{eqnarray}
where $M_S$ is the isoscalar mass, and $g$ is related to its partial
width into the Goldstone fields,
\begin{equation}
\Gamma_S\ =\ {3 g^2 M^3_S\over 32 \pi v^2}\ .
\label{threeseventeen}
\end{equation}

To this order, the Lagrangian (\ref{lscalar})\ is the most general
chirally-symmetric coupling of a spin-zero isoscalar resonance to the
fields $w^a$.  It contains two free parameters, which can be traded for
the mass and the width of the $S$.  For $g = 1$, the $S$ reduces
to an ordinary Higgs boson.  For $g \ne 1$, however, the $S$ is {\it not}
a typical Higgs boson.  It is simply an isoscalar resonance of arbitrary
mass and width.  In either case, one must be sure to check
that the scattering amplitudes are unitary up to the energy of
interest.

The tree-level scattering amplitude is easy to construct.  It has
two terms.  The first is a direct four-Goldstone coupling which
ensures that the scattering amplitude satisfies the Low-Energy
Theorems (LET) \cite{loweth,CGG}.
The second contains the contributions from the isoscalar resonance.
Taken together, they give the full scattering amplitude,
%
%\footnote{We
%use the Breit--Wigner prescription to handle the $s$--channel
%resonance. Our criterion is that all the partial waves must respect
%the unitarity condition up to 2 TeV except near the resonance; the
%slight unitarity violation near the resonance is due to the
%perturbative expansion of the width \cite{rwwll,ny}.}
%
\begin{equation}
A(s,t,u)\  = \ {s\over v^2}\ -\ \bigg({g^2 s^2 \over v^2}\bigg)
{1\over s - M_S^2 + i M_S \Gamma_S \theta (s)}\ .
\label{threeeighteen}
\end{equation}

\subsection{Spin-one, Isospin-one Resonances}

This example provides a relatively model-independent description of
the techni-rho resonance that arises in most technicolor theories.
As above, one can use the techniques of nonlinear realizations to
construct the most general coupling consistent with chiral
symmetry \cite{ccwz,BESS,bess,besslimit}.

To find the techni-rho Lagrangian, we first parameterize the Goldstone
fields $w^a$ in a slightly different way,
\begin{equation}
\xi\ =\ \exp(i \vec{\sigma}\cdot\vec{w}/2v)\ ,
\label{xidef}
\end{equation}
so $\Sigma = \xi^2$.  We then represent an $SU(2)_L \times SU(2)_R$
transformation on the field $\xi$ as follows:
\begin{equation}
\xi\ \to\ \xi^\prime\ \equiv
\ L\, \xi\, U^\dagger\ =\ U\, \xi\, R^\dagger\ .
\label{threetwenty}
\end{equation}
Here $L,\ R$ and $U$ are $SU(2)$ group elements, and $U$ is a
(nonlinear) function of $L,\ R$ and $w^a$, chosen to restore
$\xi^\prime$ to the form (\ref{xidef}).  Note that when $L = R$, $U = L
= R$ and the transformation linearizes.  This simply says that the
$w^a$ transform as a triplet under the diagonal $SU(2)$.

Given these transformations one can construct the following currents,
\begin{eqnarray}
J_{\mu L}\ & = &\ \xi^\dagger \partial_\mu
\xi \ \to\ U J_{\mu L}
U^\dagger  \ +\ U \partial_\mu U^\dagger\ , \nonumber \\
J_{\mu R}\ & = &\ \xi \partial_\mu
\xi^\dagger \ \to\ U J_{\mu R}
U^\dagger  \ +\ U \partial_\mu U^\dagger\ .
\label{jtrans}
\end{eqnarray}
The currents $J_{\mu L}$ and $J_{\mu R}$ transform as gauge fields under
transformations in the diagonal $SU(2)$.  As above, the
transformations linearize when $L=R=U$.

The transformations (\ref{jtrans})\ inspire us to choose the techni-rho
transformation as follows,
\begin{equation}
V_\mu \ \to\ U V_\mu U^\dagger
 \ +\ i\gv^{-1}\, U \partial_\mu U^\dagger\ .
\label{trhotrans}
\end{equation}
In this expression, $V_\mu = V_\mu^a \sigma^a$, and $\gv$
is the techni-rho
coupling constant.  When $L=R=U$, Eq.~(\ref{trhotrans})\ implies
that the techni-rho transforms as an isotriplet of weak isospin.

Using these transformations, it is easy to construct the most general
Lagrangian consistent with chiral symmetry.  We first write down the
currents
\begin{eqnarray}
\A_\mu\ & = &\  J_{\mu L} \ -\ J_{\mu R}\ , \nonumber \\
\V_\mu\ & = &\  J_{\mu L} \ +\ J_{\mu R}\ +\ 2i\gv V_\mu \ ,
\label{threetwentyfive}
\end{eqnarray}
which transform as follows under an arbitrary chiral transformation,
\begin{eqnarray}
\A_\mu\ &\to&\ U \A_\mu U^\dagger\ , \nonumber \\
\V_\mu\ &\to&\ U \V_\mu U^\dagger \ .
\label{threetwentysix}
\end{eqnarray}
Under parity (which exchanges $J_L$ with $J_R$ and leaves $V$
invariant), $\V$ is invariant, while $\A$ changes sign.  If we make
the additional assumption that the underlying dynamics conserve
parity, we are led to the following Lagrangian,
\begin{equation}{
\L_{{\rm Vector}}\ =
%\ -\ {v^2 \over 4}\, \Tr \partial_\mu \Sigma^\dagger\partial^\mu \Sigma
\ -\ {1\over 4}\, V^a_{\mu\nu} V^{a\mu\nu}
\ -\ {1\over4}\,v^2\,\Tr \A_\mu \A^\mu
 \ -\ {1\over4}\,a v^2\,\Tr \V_\mu \V^\mu\ +\ \ldots \ ,}
\label{threetwentyseven}
\end{equation}
where $V^a_{\mu\nu}$ is the (nonabelian) field-strength for the
vector field $V^a_\mu$.  The dots in this equation denote terms with
more derivatives.  Up to a possible field redefinition, this is the
most general coupling of a techni-rho resonance to the Goldstone
bosons, consistent with $SU(2)_L \times SU(2)_R$ symmetry.

In this Lagrangian, the parameter $v$ is fixed as before.  The
parameters $\gv$ and $a$, however, are free.  One combination is
determined by the mass of the techni-rho,
\begin{equation}
M^2_V\ =\ a {\gv}^2 v^2\ ,
\label{threetwentyeight}
\end{equation}
and another by its width into techni-pions (\ie\ Goldstone bosons),
\begin{equation}
\Gamma_V \ =\ {a M_V^3 \over 192 \pi v^2} \ .
\label{threetwentynine}
\end{equation}
The couplings described in this section are related to those of the
BESS model \cite{BESS,bess,besslimit}.
In particular $a$ corresponds to the BESS parameter $\alpha$,
and
\begin{equation}
\gv^2 = {M_V^5 \over 192 \pi v^4 \Gamma_V} =
{g^{\prime\prime 2}_{\rm BESS} \over 2}
\label{gvbess}
\end{equation}
Using this equation we can interpret the bounds on the BESS model
coupling $g^{\prime\prime}$ as bounds on the ratio $M^5_V/\Gamma_V$.

The couplings to fermions are also determined by the formalism of nonlinear
realizations, and it is possible to introduce direct couplings between the
vector resonance and ordinary fermions. These additional couplings
($b$ and $b^\prime$ in the BESS model) are not studied in this chapter.

Once again, the Goldstone-boson scattering amplitude is easy to
compute. It contains a direct four--Goldstone-boson coupling, as well as the
isovector resonance.  One finds
\begin{eqnarray}
A(s,t,u)\ & = &\ {s\over 4 v^2}\,\bigg(4 - 3\,a\bigg)
\ +\ { a M^2_V \over 4 v^2}\,\bigg[ {u-s \over t - M^2_V + i M_V \Gamma_V
\theta (t)}\nonumber \\
&& \qquad\qquad  +\ {t-s \over u - M^2_V + i M_V \Gamma_V \theta (u)}
\bigg]\ .
\label{vecamp}
\end{eqnarray}

\subsection{The Hidden Symmetry Breaking Sector}

\subsubsection{Introduction}

The symmetry breaking scenario discussed in this section has been
termed a ``Hidden Symmetry Breaking Sector'' \cite{hidden}.

As we have seen, there are various possibilities for the properties
of the quanta at the 1 TeV scale.  In the weakly coupled one-doublet
Higgs model, the new physics is a light Higgs boson.  In minimal
technicolor \cite{techni}, the exchange of the technirho and other
particles unitarizes gauge boson scattering just as the analogous
particles unitarize $\pi\pi$ scattering amplitudes in QCD.

It is frequently assumed that these two types of behavior for
elastic $W$ and $Z$ scattering are generic (see, for example
\cite{Chan,NoLoseII}).  If the symmetry breaking sector is weakly
coupled, the growth of the $W_L W_L$ scattering amplitudes is cut
off by narrow resonances (like a light Higgs boson) at a mass scale
well below a TeV. For strongly coupled theories, it is assumed that
the amplitudes saturate unitarity and that there are broad
resonances in the TeV region where the strong interaction sets in.

There is another possibility: if the electroweak symmetry breaking
sector has a large number of particles, the {\it elastic} $W$ and
$Z$ scattering amplitudes can be small and structureless,
i.e. lacking any discernible resonances.  Nonetheless, the theory
can be strongly-interacting and the {\it total} $W$ and $Z$ cross
sections large: most of the cross section is for the production of
particles other than the $W$ or $Z$. In such a model, discovering
the electroweak symmetry breaking sector depends on the observation of
the other particles and the ability to associate them with symmetry
breaking.  Physicists must keep an open mind about the experimental
signatures of the electroweak symmetry breaking sector because
discovery of electroweak symmetry breaking may not rely solely on
two-gauge-boson final states.

\subsubsection{The $O(N)$ Model}

This scenario may be illustrated by considering a toy model of the
electroweak symmetry breaking sector based on an $O(N)$ linear sigma
model.  This model is particularly interesting since it can be
solved (even for strong coupling) in the limit of large $N$
\cite{Coleman}.  One constructs a model with both exact Goldstone
bosons (which will represent the longitudinal components of the $W$
and $Z$) and pseudo-Goldstone bosons. To this end let $N=j+n$ and
consider the Lagrangian density
\begin{eqnarray}
{\cal L} = \half (\partial \vec{\phi})^2 +
\half (\partial \vec{\psi})^2 - \half \mu_{0\phi}^2 \vec{\phi}^2 - \half
\mu_{0\psi}^2 \vec{\psi}^2 - {\lambda_0 \over 8 N} {(\vec{\phi}^2 +
\vec{\psi}^2)}^2 ,
\end{eqnarray}
where $\vec{\phi}$ and $\vec{\psi}$ are $j$- and $n$-component real
vector fields. This theory has an approximate $O(j+n)$ symmetry
which is softly broken to $O(j) \times O(n)$ so long as
$\mu_{0\phi}^2 \neq \mu_{0\psi}^2$. If $\mu_{0\phi}^2$ is negative
and less than $\mu_{0\psi}^2$, one of the components of $\vec{\phi}$
gets a vacuum expectation value (VEV), breaking the approximate
$O(N)$ symmetry to $O(N-1)$. With this choice of parameters, the
exact $O(j)$ symmetry is broken to $O(j-1)$ and the theory has $j-1$
massless Goldstone bosons and one massive Higgs boson. The $O(n)$
symmetry is unbroken, and there are $n$ degenerate pseudo-Goldstone
bosons of mass $m_\psi$ ($m^2_\psi = \mu^2_{0\psi} -
\mu^2_{0\phi}$).  This section considers this model in the limit
that $j,n \to \infty$ with $j/n$ held fixed\footnote{For the
complete details of the solution of this model, see
\cite{hidden} and \cite{scalars}.}.

The scalar sector of the standard one-doublet Higgs model has a
global $O(4) \approx SU(2) \times SU(2)$ symmetry, where the 4 of
$O(4)$ transforms as one complex scalar doublet of the
$SU(2)_W\times U(1)_Y$ electroweak gauge interactions. This symmetry
is enlarged in the $O(N)$ model: the spin-0 weak isosinglet
scattering amplitude of longitudinal gauge bosons is modeled by the
spin-0 $O(j)$ singlet scattering of the Goldstone bosons in the
$O(j+n)$ model solved in the large $j$ and $n$ limit.  Of course,
$j=4$ is not particularly large. Nonetheless, the resulting model
will have all of the qualitative features needed, and the Goldstone
boson scattering amplitudes will be unitary (to the appropriate
order in $1/j$ and $1/n$).  Thus this theory can be used to
investigate the scattering of Goldstone bosons at moderate to strong
coupling \cite{Einhorn}.  No assumptions need be made about the
embedding of $SU(2)_W \times U(1)_Y$ in $O(n)$, {\it i.e.} the
electroweak quantum numbers of the pseudo-Goldstone bosons can be
anything; in this work it is assumed that the pseudo-Goldstone
bosons are $SU(3)$ color singlets\footnote{Gauge boson pair
production in models with colored pseudo-Goldstone bosons is
discussed in detail in \cite{gscat}.}.

One may compute Goldstone boson scattering to leading order in
$1/N$.  The amplitude $a^{ij;kl}(s,t,u)$ for the process $\phi^i
\phi^j \to \phi^k \phi^l$ is
\be
a^{ij;kl}(s,t,u) = A(s)\delta^{ij}\delta^{kl} +
A(t)\delta^{ik}\delta^{jl} +
A(u)\delta^{il}\delta^{jk}
\label{scatamp}
\ee
where
\be
A(s) = {s \over
v^2 - Ns\left({1\over\lambda(M)} +\widetilde{B}(s;m_\psi,M)\right)}
\punct[,]
\ee
and
\def\xx{\sqrt{s / (4 m_\psi^2 - s)}}
\begin{eqnarray}
\widetilde{B} &=& {n \over 32 N \pi^2}
\left\{
 1
+ {i \over \xx} \log{i - \xx \over i + \xx}
- \log{m_\psi^2 \over M^2}
\right\}\nonumber\\
&&+
{ j \over 32 N \pi^2 }
\left\{ 1 + \log{M^2 \over -s}
\right\}\punct[.]
\end{eqnarray}
Here $M$ is a renormalization point which is chosen below such that
the renormalized coupling, $\lambda(M)$, satisfies $1/\lambda(M)=0$.

Plotted in Fig.~\ref{A} is the absolute value of $a_{00}$ {\it vs.} the
center-of-mass energy for different values of $M$.  Here $j = 4$, $n
= 32$, $m_\psi = 125$ GeV, and $f = 250$ GeV.  The curves plotted
correspond to approximately $8M / m_\psi$ = 10000, 600, 200, 100,
and 60.  For the weakly coupled theory, for example the 10000 curve,
there is a light Higgs boson which decays to $\phi$'s.  When the
Higgs boson is light, its width is more or less unaffected by the
heavy $\psi$'s, and thus its properties are identical to those of
the Higgs boson of similar mass in the $O(j)$ model
\cite{Einhorn}.  As the Higgs resonance gets closer to the two $\psi$
threshold, it gets relatively narrower than it would have been were the
$\psi$'s absent.  As the theory becomes more strongly coupled still, the
resonance gets heavier and broader. Eventually, for small enough $M$, the
imaginary part of the location of the pole is so great that there is no
discernible resonance in $a_{00}$.

\begin{figure}[htb]
% Han's change:
%%%\centerline{\epsfxsize=4.5in\epsfbox{fig1.eps}}
\vspace{2.0in}
\caption{\label{A}
The absolute value of the $\phi\phi\to\phi\phi$ scattering
amplitude vs. CM energy for different values of $M$. Here $j = 4$, $n = 32$,
$m_\psi = 125$ GeV, and $f = 250$ GeV. The curves correspond to roughly $8M /
m_\psi \sim $ 10000, 600, 200, 100, and 60.  The curve with the leftmost bump
is 10000, and the low nearly structureless curve is $8 M / m_\psi \sim
60$. For comparison, the dashed line shows the scattering amplitude in the
limit $m_\psi \to \infty$ with $M$ adjusted to produce a Higgs resonance at
approximately 500 GeV.}
\end{figure}

When the Higgs resonance is heavier than twice $m_\psi$, it no longer decays
exclusively to $\phi$'s, and thus the absolute value of the amplitude for
elastic $\phi\phi$ scattering never gets anywhere near 1.  Probability is
leaking out of this channel into that for the production of pairs of $\psi$'s.
For comparison, the dashed line shows the scattering amplitude in the limit
$m_\psi \to \infty$ with $M$ adjusted to produce a Higgs resonance at
approximately 500 GeV.

In the gauged model, the $\phi$'s are eaten by the gauge bosons, and become
their longitudinal components.  Therefore, $\phi\phi$ final states correspond
to two-gauge-boson events.  In this toy model the Higgs resonance may be light
but so broad that at no energy is the number of $WW$ or $ZZ$ events large;
discovering the Higgs boson depends on its observation in the $\psi\psi$
channel.  Depending on how the $\psi$'s decay, this may be easy or hard.
Nonetheless, it is clear that an experiment looking for electroweak symmetry
breaking may not be able to rely exclusively on the two gauge boson events.

\clearpage

\subsection{Current Experimental Constraints}

\subsubsection{Bounds on $L_{1,2}$ from LEP data}

One can use the effective Lagrangian to study processes at low
energies, where the longitudinal degrees of freedom are not dominant.
For that purpose it is necessary to consider the complete effective
Lagrangian for the electroweak gauge bosons. In this way, one can place
indirect bounds on the couplings $L_{1,2}$ through their loop contributions
to already measured processes. For example,
the couplings $L_{1,2}$ enter the one-loop
calculation of the $Z \to f \overline{f}$
width. With some assumptions it is possible to use these partial
widths to place the $90\%$ confidence level bounds \cite{dawvallep}:
\begin{equation}
-28 \leq L_1 + {5 \over 2} L_2 \leq 26.
\label{combineonetwo}
\end{equation}

\subsubsection{Present limits on BESS Model Parameters}

LEP data including the total $Z$ width, the hadronic
and the leptonic partial $Z$ widths, the leptonic and bottom forward-backward
asymmetries, the $\tau$-polarization together with the cesium atomic parity
violation and the ratio $M_W / M_Z$, have been analyzed in terms of the
parameters $\varepsilon_i$ \cite{alta}.

The BESS contribution\footnote{The results on the BESS model
in this and the
latter sections were provided
by R. Casalbuoni, P. Chiappetta, A. Deandrea, S. De Curtis,
D. Dominici, R. Gatto \cite{besstev}.}
to these parameters is \cite{besstev}:
\be
\varepsilon_1 =  \varepsilon_2 = 0 \nonumber ; \ \
\varepsilon_3 =  \left( \frac{g}{g''} \right)^2 - \frac{b}{2},
\ee
where $b$ is a parameter that characterizes the possible
direct coupling between the fermions and the new vector bosons.

The allowed region at $90\%$ CL in the $\left(b, {g}/{g''}
\right)$ plane is shown in Figure \ref{coupl}
for a top mass value of $174\pm 17$ GeV
and in the limit $M_V \gg M_W$.
The chosen experimental value \cite{alta2}
\be
\varepsilon_3^{\exp} = (3.9 \pm 1.7) 10^{-3}
\label{altalim}
\ee
corresponds to the LEP1 data combined with UA2/CDF/D0 ones
presented at the Glasgow conference, and we have added to
Eq.~\ref{altalim} the
contribution coming from the radiative corrections \cite{alta2}
for $M_H=\Lambda=1~TeV$ and $m_{top}=174\pm 17$ GeV, which is
$\varepsilon_3^{\rm {rad.~corr.}}=(6.39^{-0.14}_{+0.20}) 10^{-3}$.
The resulting bound is shown in Figure~\ref{coupl}.

\smallskip

\noindent
\begin{figure}[htb]
\centerline{
\epsfxsize=8truecm
%\epsffile[69 263 498 700]{contr_bess4-1.ps}}
\epsffile[69 263 498 700]{contr_bess2-2.ps}}
\caption{\label{coupl}
$90\%$ C.L. contour in the plane $(b,g/\gs)$
 from the measurement of $\varepsilon_3$. The solid (dashed) line is for
$m_{top}$(GeV)= 191(157), $\Lambda=1$ TeV and $\alpha_s=0.118$.}
\end{figure}

\section{Strongly-interacting ESB Sector at Hadron Colliders}

\subsection{Vector Resonance Signals at an Upgraded Tevatron}

We first consider the detection of a signal for a vector resonance
at a possible upgrade of the Fermilab
Tevatron collider with center of mass energy 4 TeV and
an integrated luminosity of 10 fb$^{-1}$.
For a vector dominance model,  it turns out that
the $pp\to W^\pm Z+X$ process via the Drell-Yan (DY)
mechanism is most useful.
Due to the lower energy of  such a  collider, the vector boson fusion
processes are much less significant.

In Ref.~\cite{besstev} examples with different choices of parameters
$M_V$, $b$ and $g''$ are studied. Figure \ref{figsix}  shows
the prediction for invariant $W^+Z$ mass distribution
for $M_V=600$ GeV, $\gs=13$ and $b=0.01$ to which corresponds
a width $\Gamma_V=0.9$ GeV.
The signal is doubled by adding the $W^-Z$ channel  final state.
To avoid the QCD jet background, for most of the discussions
in this report,
only the ``gold-plated'' channels of pure leptonic decays,
such as $W^\pm Z \rightarrow l^\pm \nu l^+ l^-$ are considered.
After multiplying by the appropriate leptonic branching ratios,
one is left with a statistically significant signal with
$S/\sqrt B \simeq  27/3=9$.

\begin{figure}[htb]
\centerline{
\epsfxsize=8truecm
\epsffile[45 214 551 654]{mwz3-1.ps}}
\smallskip
\smallskip
\smallskip
\caption{\label{figsix}
Invariant mass distribution of the $W^+Z$ pairs produced
per year at Tevatron Upgrade for $M_V=$600 GeV, $\gs=13$ and $b=0.01$.
The applied cuts are
$(p_T)_Z>$180 GeV and $M_{WZ}>$500 GeV. The lower, higher
histograms refer the background (296 events),
background plus $q\bar q$ annihilation signal
(912 events).}
\end{figure}

The above results depend significantly on the values
of  model parameters $b$ and $\gs$. For the same $V$ mass the
case corresponding to the choice $\gs=20$ and $b=0.016$
leads to roughly five times more events. Increasing the mass
to  800 GeV reduces the signal by roughly a factor of five.
In a definite region of the parameters $(b,g/\gs)$, the discovery limit of the
Tevatron Upgrade can reach masses $M_V\sim 1$ TeV \cite{besstev}.

\subsection{Complementarity of Vector Resonance and \hfill\break
Non-resonant $WW$ Scattering at LHC}

%
%\subsubsection{BESS at LHC}

The DY mechanism is found to be in general
dominant at the LHC with respect to the fusion mechanism.
It seems that a mass discovery limit for charged vector resonances
around 2 TeV can be achieved at the LHC for a large domain of the BESS
parameter space \cite{BESSssc,pseudo}.
Nevertheless there are still some parameter values which lead, even
for light $M_V$ masses, to too small number of events to be discovered.

%
%Chanowitz & Kilgore
%

Using a QCD-like chiral Lagrangian with a dominant ``$\rho$'' meson to
study strong scattering in  the $WZ$ and $W^+W^+$ channels,
Ref.~\cite{mcwk} discussed a complementary relationship
between
the resonant ``$\rho$'' signal, best observed in the $WZ$ channel, and
nonresonant scattering in the $W^+W^+$ channel.  Namely,
for smaller $m_{\rho}$ the resonant $\rho \rightarrow WZ$ signal is large while
the nonresonant $W^+W^+$ signal is suppressed. For very large $m_{\rho}$ the
resonant $WZ$ signal is unobservable but nonresonant $W^+W^+$ scattering is
large, approaching the K-matrix unitarization of the low energy theorem as
$m_{\rho} \rightarrow \infty$
\footnote{In their calculation, the mass $m_{\rho}$ is a free parameter
an the width $\Gamma_{\rho}$ is fixed by scaling from QCD.}.

For $WZ$ channel, the major
backgrounds considered are the continuum $q \bar q \rightarrow WZ$ and the
electroweak $q q \rightarrow q q WZ$ of ${\cal O}(\alpha^4)$; in addition
there are top quark induced backgrounds which were not included, but which are
expected to be largely eliminated by the cuts imposed (see next section).

For the $W^\pm W^\pm$ channel, the dominant backgrounds are the ${\cal
O}(\alpha_W^2)$ \cite{dv2} and ${\cal O}(\alpha_W \alpha_S)$ \cite{mcmg-dv1}
amplitudes for $qq \to qqWW$.  The former, like the analogous $WZ$ background,
is computed in the standard model with a light Higgs boson.  Other
backgrounds, from $W^+W^-$ with lepton charge mismeasured and from $\overline
tt$ production are detector dependent and are therefore not included, but are
expected to be controllable \cite{sdctdr} (see next section).

For the $WZ$ channel the lepton rapidity was required to be $y_{l} < 2$.  The
cuts on the $Z$ transverse momentum, $p_{TZ} > p_{TZ}^{MIN}$, and on the
azimuthal angles $\phi_{ll}$ between the leptons from the $Z$ and the charged
lepton from the $W$, $\hbox{cos}\phi_{ll} < (\hbox{cos}\phi_{ll})^{MAX}$, were
optimized for each choice of $m_{\rho}$ and for each collider energy. A
central jet veto was also imposed on events with one or more hadronic jets
with rapidity $y_j < 3$ and transverse momentum $p_{Tj} > 60$ GeV.  In
addition, a detector efficiency of 80\% for the process was assumed
\cite{sdctdr}.

For the like-charged $W^+W^+$ and $W^-W^-$ channels, there were cuts on the
lepton transverse momentum $p_{Tl}$, on the azimuthal angle between the two
leptons $\phi_{ll}$, and a veto on events with central jets as defined above.
Lastly, an 85\% detection efficiency for a single isolated lepton was used.

The observability criteria are \cite{mcwk}
\begin{equation}
\sigma^{\uparrow}   =  S/\sqrt{B}  \ge  5 ~~~\hbox{,}~~~
\sigma^{\downarrow}   =  S/\sqrt{S+B}  \ge  3 ~~~\hbox{and}~~~
S \ge B,
\label{CRIT}
\end{equation}
where $S$ and $B$ are the number of signal and background events, and
$\sigma^{\uparrow}$ and $\sigma^{\downarrow}$ are respectively the number of
standard deviations for the background to fluctuate
up to give a false signal or for the signal plus background to
fluctuate down to the level of the background alone.
In addition, one must require $S \ge B$ so that the signal is
unambiguous despite the
systematic uncertainty in the size of the backgrounds,
expected to be known to within $ \leq \pm 30 \%$ after ``calibration'' studies
at the LHC.

The statistical criteria apply to the detected events, i.e., {\em after}
efficiency corrections are applied.  For example, assuming 85\% detection
efficiency for a single isolated lepton , the criterion for the $W^+W^+$
signal applied to the uncorrected yields is $\sigma^{\uparrow}>6$ and
$\sigma^{\downarrow}>3.5$.

Complementarity of the resonant and nonresonant channels is evident in the
inverse relationship between ${\cal L}_{MIN}(WZ)$ and ${\cal L}_{MIN}(WW)$
in Table~\ref{xxytable}. For the like-sign $WW$ channel the
biggest signal occurs for the heaviest $\rho$
meson, $m_{\rho}=4$ TeV, with a signal meeting the significance criterion for
77 fb$^{-1}$. The $WZ$ signal is smallest for that case, and there is
no cut for which Eq.~(\ref{CRIT}) is satisfied, indicated by ``NS''
(no signal) in the table. For $m_{\rho}=1.78$ TeV the
like-sign $WW$ signal is smaller but the resonant $WZ$ signal is much larger
and satisfies the
criterion with only 44 fb$^{-1}$. The worst case among the four models
is at $m_{\rho}=2.52$ TeV.
Because of the $l^+l^-l^+\nu$ background, the
optimum cut for that model, still in the like-sign $WW$ channel, requires
105 fb$^{-1}$ to meet the criterion.
%, compared to 63 fb$^{-1}$ in \cite{mcwk}.

\begin{table}[htbp]
\centering
\caption{\label{xxytable}%
Minimum luminosity to satisfy significance criterion for $W^+Z + W^-Z$
and $W^+W^+ + W^-W^-$ scattering. For the like-sign $WW$ channel, the
table also shows
the optimum cut on the like-sign leptons that gives ${\cal L}_{MIN}(WW)$,
the corresponding number of signal and background events per 100 fb$^{-1}$, and
the composition of the background for
the optimum cut. A central jet veto is included as specified in the text.
Rejection of all events for which the third lepton
falls within its acceptance region is assumed.
}
\bigskip
\begin{tabular}{ccccc}
$m_{\rho}$(TeV) & 1.78 & 2.06&2.52&4.0\cr
\hline
\hline
&&&&\cr
${\cal{L}}_{MIN}(WZ)\ ({\rm{fb}}^{-1})$ &44 &98 &323&NS\cr
${\cal{L}}_{MIN}(WW)\ ({\rm{fb}}^{-1})$&142&123&105&77\cr
&&&&\cr
\hline
&&&&\cr
$WW$ Cut&&&&\cr
&&&&\cr
$\eta^{MAX}(l)$&1.5&1.5&1.5&2.0\cr
$p^{MIN}_{T}(l)$ (GeV)&130.&130.&130.&130.\cr
$[\cos\phi(ll)]^{MAX}$&$-0.72$&$-0.80$&$-0.80$&$-0.90$\cr
&&&&\cr
\hline
&&&&\cr
$WW$ Sig/Bkgd&&&&\cr
&&&&\cr
(events per $100\ {\rm{fb}}^{-1})$&12.7/6.0&14.1/5.8&15.9/5.8&22.4/8.9\cr
&&&&\cr
\hline
&&&&\cr
$WW$ Backgrounds (\%)&&&&\cr
&&&&\cr
$\overline l l \overline l \nu_{l}$&47&49&49&61\cr
$O(\alpha^2_W)$&47&46&46&33\cr
$O(\alpha_W\alpha_S)$&6&6&6&6\cr
\end{tabular}
\end{table}

\subsection{All Channel Comparison for $WW$ Scattering at  LHC}

In Ref. \cite{baggeretali}, systematic comparisons for all
$WW$ scattering channels and comprehensive background calculations
were carried out.  The final state modes considered were the following:
\[
\begin{array}{c}
ZZ\to \ell^+\ell^-\ell^+\ell^-, \qquad ZZ\to
\ell^+\ell^-\nu\overline\nu, \qquad
W^{\pm} Z\to \ell^\pm\nu\ell^+\ell^-,  \\
W^+W^-\to \ell^+\nu\ell^-\overline\nu,
 \qquad W^{\pm}W^{\pm} \to \ell^\pm\nu\ell^\pm\nu .
\end{array}
\]
%
%\begin{enumerate}
%
%\item
%the ``SM'' with $m_H=1~{\rm TeV}$;
%
%\item
%the ``Scalar'' model with a spin-0, isospin-0
%chirally coupled resonance with mass of $1~{\rm TeV}$
%and width $350~{\rm GeV}$;
%
%\item
%an ``$O(2N)$'' model with $N=2$ and
%an amplitude having a pole at $s=[m-i\Gamma/2]^2$ with
%$m=0.8~{\rm TeV}$ and $\Gamma=600~{\rm GeV}$;
%
%\item
%a ``Vector'' model with a spin-1, isospin-1
%chirally coupled resonance;  the mass-width combinations are
%($M_V,\Gamma_V$)=($1~{\rm TeV},5.7~{\rm GeV}$) and
%($2.5~{\rm TeV},520~{\rm GeV}$);
%
%\item
%the non-resonant ``LET-CG'' model of Ref. \cite{changail} in which the low
%energy theorem (LET) amplitude is used
%and unitarity saturation is assumed once  the partial waves
%reach the unitarity bound;
%
%\item
%the non-resonant  ``LET-K'' model in which the low energy theorem amplitude is
%used and unitarization of the partial waves is achieved via the K-matrix
%technique;
%
%\item
%the ``Delay-K'' model in which one-loop correction terms
%to the LET amplitude are chosen so as to delay the onset
%of unitarity violation to energies beyond $2~{\rm TeV}$, and K-matrix
%unitarization
%is employed to ensure unitarity beyond this point.
%
%\end{enumerate}
%
%\noindent
%These models will be referenced using a short-hand notation: SM, Scalar,
%$O(2N)$, Vec1.0 (Vec2.5), LET-CG, LET-K, and Delay-K, respectively.
%They are fully described in Ref.~\cite{baggeretali}.
%

The signal events from strong $W_LW_L$  scattering processes
with  $W_LW_L$ masses of  order $\cal O$(1 TeV) are characterized
by several unique
features \cite{RPHstrong_Barger,bckttbar,wpwp}:
\begin{enumerate}

\item[(i)] very energetic leptons in the central
(low rapidity) region coming from the two fast $W_L$'s;
a very back-to-back structure for the leptons,
such that $\cos\phi_{\ell\ell}$ is very near $-1$
($\phi_{ll}$ is the azimuthal angle between a lepton $\ell_1$ from
one $W_L$ and a lepton $\ell_2$ from the other $W_L$) and $\Delta
p_T(\ell\ell)\equiv
|{\bf p_T}(\ell_1) - {\bf p_T}(\ell_2)|$
is very large; and, of course, high values for the invariant
mass $M(\ell\ell)^2=(p_{\ell_1}+p_{\ell_2})^2$
constructed from two such leptons;

\item[(ii)] low hadronic jet activity in the central region;

\item[(iii)] highly energetic,
low-$p_T$ (i.e.\ high rapidity, $|y|$) forward-backward
`spectator' jets \cite{RPHlnujj_tagth}.
\end{enumerate}

\noindent
By imposing stringent leptonic cuts, vetoing central hard jets, and
tagging energetic forward spectator jets to
single out these features of the $W_LW_L$ signal,
the otherwise very large backgrounds can be suppressed dramatically.
%and a viable signal for strong $W_LW_L$ scattering established.
In particular, the most difficult ``irreducible''  electroweak backgrounds from
$W_TW_T$ and $W_TW_L$ scattering have spectator jets that are more central
and yield leptons that are not  so back-to-back. The central-jet-vetoing,
$\cos\phi_{\ell\ell}\sim -1$ and large $\Delta p_T(\ell\ell)$ cuts prove to be
effective against those backgrounds.
The resultant optimized cuts are listed in Table~\ref{cutstable}.

The procedure for
determining the efficiency with which the $W_LW_L$ signal is retained
for a given set of cuts for each of the models is as follows.
First, perform
an exact calculation employing full SM matrix elements for
$q\overline q' \to q\overline q' WW\to q\overline q'+4~{\rm
leptons}$
for the cases of a  1 TeV Higgs boson and a light Higgs boson (0.1 TeV).
Regardless of the strongly-interacting $W_LW_L$ sector model,
the light Higgs boson results should accurately represent the perturbative
SM ``irreducible'' backgrounds from $qq'  \to qq'  W_TW_T (W_TW_L)$
processes that are present in the absence of strong $W_LW_L$ scattering.
This background is inevitably present due to the inability of experimentally
fixing the polarizations of the final state $W$'s on an event by event basis.
Then, the $W_LW_L$ signal for a 1 TeV Higgs boson
is defined as the enhancement over the SM prediction
with a light Higgs boson (say, 0.1 TeV):
\begin{equation}
\sigma(W_LW_L~{\rm signal})\equiv \sigma({\rm SM}~m_H=1~{\rm TeV})-
     \sigma({\rm SM}~m_H=0.1~{\rm TeV})\,.
\label{signaldef}
\end{equation}
The jet-cut efficiencies for a 1 TeV Higgs boson signal, which
are later applied to other models,  are determined based
on the signal definition of Eq.~(\ref{signaldef}).  The SM
results for cross sections after imposing the cuts in Table~\ref{cutstable},
and the resulting cut efficiencies  for  a 1 TeV Higgs boson signal
and all the other SM backgrounds are given in Table~\ref{smlhctable}.
Details regarding these background computations can be   found in
Ref.~\cite{baggeretali}.

\begin{table}[htbp]
\centering
\caption{\label{cutstable}
Leptonic cuts, single-tagging and central-vetoing
cuts on jets  for generic $W_LW_L$ fusion processes
at the LHC energy,  by final state  mode.}
\bigskip
\begin{tabular}{lcc}
\hline\hline
$Z Z(4\ell)$ & leptonic cuts & jet cuts \\
\hline
 \ &
 $\vert  y({\ell}) \vert  < 2.5 $  &
 $E(j_{tag}) > 0.8~{\rm TeV}$   \\
 \ &
 $p_T(\ell) > 40~{\rm GeV}$  &
 $3.0 < \vert y(j_{tag}) \vert < 5.0$   \\
 \ &
 $p_T(Z) > {1\over4} \sqrt{M^2({ZZ}) - 4 M^2_Z}$  &
 $p_T(j_{tag}) > 40~{\rm GeV}$   \\
 \ &
 $M({ZZ}) > 500~{\rm GeV}$  & no veto \\
\hline
$Z Z(\ell\ell\nu\nu)$ & leptonic cuts & jet cuts \\
\hline
 \ &
 $\vert  y({\ell}) \vert  < 2.5 $  &
 $E(j_{tag}) > 0.8~{\rm TeV}$   \\
 \ &
 $p_T(\ell) > 40~{\rm GeV}$  &
 $3.0 < \vert y(j_{tag}) \vert < 5.0$   \\
 \ &
 $p_T^{\rm miss} > 250~{\rm GeV}$ &
 $p_T(j_{tag}) > 40~{\rm GeV}$   \\
 \ &
 $M_T > 500~{\rm GeV}$  &
 $p_T(j_{veto}) > 60~{\rm GeV}$ \\
 \ &
 $p_T{(\ell\ell)}>M_T/4$ &
 $ \vert  y(j_{veto}) \vert  < 3.0$  \\
\hline
$W^+W^-$ & leptonic cuts & jet cuts \\
\hline
 \ &
 $\vert y({\ell}) \vert < 2.0 $  &
 $E(j_{tag}) > 0.8~{\rm TeV}$  \\
 \ &
 $p_T(\ell) > 100~{\rm GeV}$  &
 $3.0 < \vert y(j_{tag}) \vert < 5.0$  \\
 \ &
 $\Delta p_T({\ell\ell}) > 440~{\rm GeV}$  &
 $p_T(j_{tag}) > 40\ {\rm GeV}$  \\
 \ &
 $\cos\phi_{\ell\ell} < -0.8$  &
 $p_T(j_{veto}) > 30~{\rm GeV}$  \\
 \ &
 $M({\ell\ell}) > 250~{\rm GeV}$  &
 $ \vert  y(j_{veto}) \vert  < 3.0$  \\
\hline
$W^{\pm} Z$ & leptonic cuts & jet cuts\\
\hline
 \ &
 $\vert  y({\ell}) \vert  < 2.5 $  &
 $E(j_{tag}) > 0.8~{\rm TeV}$  \\
 \ &
 $p_T(\ell) > 40~{\rm GeV}$  &
 $3.0 < \vert y(j_{tag}) \vert < 5.0$  \\
 \ &
 $ p_{T}^{\rm miss} >  50~{\rm GeV}$  &
 $p_T(j_{tag}) > 40~{\rm GeV}$   \\
 \ &
 $p_T(Z) > {1\over4} M_T $ &
 $p_T(j_{veto}) > 60~{\rm GeV}$  \\
 \ &
 $M_T > 500\ {\rm GeV}$ &
 $ \vert  y(j_{veto}) \vert  < 3.0$  \\
\hline
$W^{\pm}W^{\pm}$ & leptonic cuts & jet cuts \\
\hline
 \ &
 $\vert  y({\ell}) \vert  < 2.0 $   &
   \\
 \ &
 $p_T(\ell) > 70~{\rm GeV}$   &
  $3.0 < \vert y(j_{tag}) \vert < 5.0$  \\
 \ &
 $\Delta p_T(\ell\ell) > 200~{\rm GeV}$   &
  $p_T(j_{tag}) > 40~{\rm GeV}$   \\
 \ &
 $\cos\phi_{\ell\ell} < -0.8$  &
 $p_T(j_{veto}) > 60~{\rm GeV}$ \\
 \ &
 $M(\ell\ell) > 250~{\rm GeV}$ &
 $\vert y(j_{veto}) \vert < 3.0$  \\
\hline\hline
\end{tabular}
\end{table}

\begin{table}[htbp]
\centering
\caption{\label{smlhctable}%
Standard Model cross sections (in fb) for electroweak
processes $q\bar q' \to q\bar q' WW$
for $m_H=1$ TeV and  $0.1$ TeV,  and for the continuum $WW$
production at ${\cal O}(\alpha_s)$ and other
backgrounds, with $\protect\sqrt{s} = 14$ TeV and  $m_t = 175$ GeV. }
\bigskip
\begin{tabular}{cccc}
\hline\hline
%$Z Z(4\ell) $ & leptonic cuts only & $+$ veto only / veto eff. &
%$+$ veto $+$ tag / tag eff. \\
$Z Z(4\ell) $ & leptonic & $+$ veto / veto eff. &
$+$ veto $+$ tag / tag eff. \\
\hline
 EW($m_H=1.0~{\rm TeV}$) & 0.12  & - & 0.045 / 39\% \\
 EW($m_H=0.1~{\rm TeV}$) & 0.019  & - & 0.004 / 19\% \\
  Continuum $ZZ$           & 0.42  & - & 0.003 / 0.6\% \\
 $Z_LZ_L$ signal     & 0.096  & - & 0.041 / 43\% \\
\hline
%$Z Z(2\ell2\nu) $ & leptonic cuts only & $+$ veto only / veto eff. &
%$+$ veto $+$ tag / tag eff. \\
$Z Z(2\ell2\nu) $ & leptonic & $+$ veto / veto eff. &
$+$ veto $+$ tag / tag eff. \\
\hline
 EW($m_H=1.0~{\rm TeV}$) & 0.69  & 0.30 / 43\%  & 0.16 / 54\% \\
 EW($m_H=0.1~{\rm TeV}$) & 0.11  & 0.014 / 13\%  & 0.006 / 38\% \\
  Continuum $ZZ$          & 2.2  & 1.7 / 75\%  & 0.012 / 0.7\% \\
 $Z_LZ_L$ signal     & 0.59  & 0.29 / 49\%  & 0.16 / 55\% \\
\hline
%$W^+W^-$ & leptonic cuts only & $+$ veto only / veto eff. &
%$+$ veto $+$ tag / tag eff. \\
$W^+W^-$ & leptonic & $+$ veto / veto eff. &
$+$ veto $+$ tag / tag eff. \\
\hline
 EW($m_H=1.0~{\rm TeV}$) & 1.1 & 0.33 / 30\%  & 0.20 / 59\% \\
 EW($m_H=0.1~{\rm TeV}$) & 0.32 & 0.039 / 12\%  & 0.016 / 40\% \\
 Continuum $W^+W^-$          & 6.8 & 3.5 / 51\%  & 0.041 / 1.2\% \\
 $t\overline t+{\rm jet}$        & 59 & 0.88 / 1.5\% & 0.067 / 7.7\% \\
 $W_LW_L$ signal     & 0.80 & 0.29 / 37\%  & 0.18 / 61\% \\
\hline
%$W^{\pm} Z$ & leptonic cuts only & $+$ veto only / veto eff. &
%$+$ veto $+$ tag / tag eff. \\
$W^{\pm} Z$ & leptonic & $+$ veto / veto eff. &
$+$ veto $+$ tag / tag eff. \\
\hline
 EW($m_H=1.0~{\rm TeV}$) & 0.32 & 0.07 / 22\% & 0.032 / 46\% \\
 EW($m_H=0.1~{\rm TeV}$) & 0.25 & 0.043 / 17\% & 0.018 / 42\% \\
 Continuum  $W^{\pm} Z$  & 3.8 & 2.2 / 56\% & 0.03 / 1.4\% \\
 $Z t\overline t+{\rm jet}$        & 0.42 & 0.008 / 2.0\% & 0.001 / 16\% \\
 $W_LZ_L$ signal     & 0.073 & 0.027 / 37\%  & 0.014 / 52\% \\
\hline
%$W^{\pm}W^{\pm}$ & leptonic cuts only & $+$ veto only / veto eff. &
%$+$ veto $+$ tag / tag eff. \\
$W^{\pm}W^{\pm}$ & leptonic & $+$ veto / veto eff. &
$+$ veto $+$ tag / tag eff. \\
\hline
 EW($m_H=1.0~{\rm TeV}$) & 0.66 & 0.15 / 23\% & 0.099 / 66\% \\
 EW($m_H=0.1~{\rm TeV}$) & 0.45 & 0.057 / 13\% & 0.034 / 60\% \\
 $g$-exchange        & 0.15 & 0.009 / 6.0\% & 0.001/ 7.7\% \\
 $Wt\overline t$         & 0.42 & 0.012 / 3.0\% & 0.001 / 13\% \\
 Continuum  $W^{\pm} Z$  & 0.15 & 0.10/ 65\% & 0.001 / 1.4\% \\
 $W_LW_L$ signal     & 0.22 & 0.093 / 43\%  & 0.066 / 70\% \\
\hline\hline
\end{tabular}
\end{table}

To compute the cross sections for $W_LW_L \to W_LW_L$ signals
in a given model beyond the SM,
one can use the Effective-$W$ Approximation (EWA) \cite{effectw}
in combination with the Equivalence Theorem \cite{equivth,lqt,changail,eqth}.
To obtain cross sections in the EWA/ET approximation that include
the jet-tagging and jet-vetoing cuts, one simply multiplies
the cross sections calculated via the EWA/ET
technique, including leptonic cuts, by the net jet-tagging and/or jet-vetoing
efficiency as obtained for the $W_LW_L$ signal
in the exact SM calculation with a 1 TeV Higgs boson
(see Table~\ref{smlhctable}).
This procedure should be fairly accurate
since the kinematics of the jets in the
signal events are determined by those of
the initial $W_L$'s that participate in the $W_LW_L$ scattering
process.  These kinematics are independent of the
strong $W_L W_L$  scattering dynamics.

\begin{figure}[htbp]
%%%\centerline{\epsfxsize=4.5in\epsfbox{scalar3-3a.eps}}
\caption{\label{sewsfigia}
Invariant mass distributions for the
``gold-plated'' leptonic final states that arise
from the processes $pp \to ZZX \to 4\ell X$,
$pp \to ZZX \to 2\ell  2\nu X$,
$pp \to W^+W^-X$, $pp \to W^\pm ZX$ and
$pp \to W^\pm W^\pm X$, for $\protect \sqrt s = 14$ TeV
and an annual LHC luminosity of 100~fb$^{-1}$, in the case of a
chirally coupled scalar with $M_S = 1$ TeV, $\Gamma_S = 350$ GeV.
The signal is plotted above the
summed background. The mass variable of
$x$-axis is in units of  GeV and the bin size is 50 GeV.}
\end{figure}
\begin{figure}
%%%\centerline{\epsfxsize=4.5in\epsfbox{vector3-3b.eps}}
\caption[]{\label{sewsfigib}
Same as Figure~\ref{sewsfigia} but for a
chirally coupled vector with $M_V = 1$ TeV, $\Gamma_V = 5.7$ GeV.}
\end{figure}
\begin{figure}
%%%\centerline{\epsfxsize=4.5in\epsfbox{vector3-3c.eps}}
\caption[]{\label{sewsfigic}
Same as Figure~\ref{sewsfigia} but for a
chirally coupled vector with $M_V = 2.5$ TeV, $\Gamma_V = 520$ GeV.}
\end{figure}
\begin{figure}
%%%\centerline{\epsfxsize=4.5in\epsfbox{letk3-3d.eps}}
\caption[]{\label{sewsfigid}
Same as Figure~\ref{sewsfigia} but for a
nonresonant model unitarized by the $K$-matrix prescription.}
\end{figure}

Table~\ref{lhcrates}\ and
Figures~\ref{sewsfigia}, \ref{sewsfigib},
\ref{sewsfigic}, and \ref{sewsfigid}
%Figures~\ref{sewsfigia,sewsfigib,sewsfigic,sewsfigid}
present the results
for $W_LW_L$ fusion signals for  various  Strongly-interacting
electroweak symmetry breaking models versus SM backgrounds.
Large excesses above SM backgrounds
are predicted in both the $ZZ(4\ell)$ and $ZZ(2\ell2\nu)$
modes in the cases of a 1 TeV SM Higgs boson,
the Scalar resonance model and the $O(2N)$ model.
Especially encouraging is the signal rate for the  $2\ell 2\nu$ mode.
Purely on the basis of the number of events,
the $W^+ W^-$ channel also shows some sensitivity to those Scalar
resonance models with a sizeable rate; actual sensitivity
in this channel is probably somewhat greater since the distribution
in the mass variable $M(\ell\ell)$ broadly peaks around $m_H/2$.
The Vec1.0 and Vec2.5 models would yield
an observable event excess in the $W^{\pm}W^{\pm}$ channel,
and, to a much lesser extent, in the $W^{\pm} Z$ channel where the
signal rates are rather low and the background level
remains difficult. However, in the $W^{\pm} Z$ channel
there might be a chance to search
for the signal peak in the $M_T$ spectrum for $M_V \sim 1$ TeV.
Finally, the LET-CG, LET-K, and Delay-K models
all yield observable excesses in the $W^{\pm}W^{\pm}$ channel.

\begin{table}[htbp]
\centering
\caption{\label{lhcrates}%
Event rates per LHC-year  for $W_LW_L$ fusion signals
for various strongly-interacting EWSB models and backgrounds,
assuming $\protect\sqrt s=14~{\rm TeV}$, an annual luminosity of $100 ~{\rm
fb}^{-1}$, and
$m_t=175~{\rm GeV}$.  Cuts are listed in Table~\protect\ref{cutstable}.
Jet-vetoing and tagging efficiencies are listed in
Table~\protect\ref{smlhctable}.}
\bigskip
\begin{tabular}{l|ccccccccc}
\hline\hline
%& Bkgd. & SM & Scalar & $O(2N)$ & Vec 1.0 & Vec 2.5 & LET-CG &
%   LET-K & Delay-K  \\
& Bkg & SM & Scal & $O(2N)$ & V1.0 & V2.5 & CG &
   LET-K & Dly-K  \\
\hline
$Z Z(4\ell)$  & 0.7 & 9 & 4.6 & 4.0 & 1.4 & 1.3 & 1.5 & 1.4 & 1.1  \\
 \hline
$Z Z(2\ell2\nu)$ & 1.8 & 29 & 17 & 14 & 4.7 & 4.4 & 5.0 & 4.5 & 3.6  \\
 \hline
$W^+W^-$ & 12 & 27 & 18 & 13 & 6.2 & 5.5 & 5.8 & 4.6 & 3.9 \\
  \hline
$W^{\pm} Z$ & 4.9 & 1.2 & 1.5 & 1.2 & 4.5 & 3.3 & 3.2 & 3.0 & 2.9 \\
  \hline
$W^{\pm}W^{\pm}$ & 3.7 & 5.6 & 7.0 & 5.8 & 12 & 11 & 13 & 13 & 8.4 \\
\hline\hline
\end{tabular}
\end{table}

{}From Table~\ref{lhcrates} one can estimate
the number of LHC years needed for
generating a signal at a  99\% Confidence Level from Poisson-type
statistics.  The results are given in Table~\ref{lhclum}.
With a few years running at the LHC,  one should be able to observe
a significant enhancement in at least one gold-plated channel, thereby
revealing the underlying EWSB physics.

\begin{table}[htbp]
\centering
\caption{\label{lhclum}%
Number of years (if $<10$) at the LHC (an annual luminosity 100 fb$^{-1}$)
required for a 99\% confidence level signal.}
\bigskip
\begin{tabular}{l|cccccccc}
\hline\hline
& \multicolumn{8}{c}{Model}\\
\cline{2-9}
%Channel& SM& Scalar& $O(2N)$& Vec1.0& Vec2.5& LET CG&
%LET K& Delay K\\
Channel& SM& Scal& $O(2N)$& V1.0& V2.5& CG&
LET-K& Dly-K\\
\hline
$ZZ(4\ell)$ &
1.0  & 2.5  & 3.2  & ~    & ~   & ~    & ~     & ~    \\
$ZZ(2\ell2\nu)$ &
0.5  & 0.75 & 1.0  & 3.7  & 4.2 & 3.5  & 4.0   & 5.7  \\
$W^+W^-$ &
0.75 & 1.5  & 2.5  & 8.5  & ~   & 9.5  & ~     & ~    \\
$W^{\pm} Z$ &
 ~   & ~    & ~    & 7.5  & ~   & ~    & ~     & ~    \\
$W^{\pm}W^{\pm}$ &
4.5  & 3.0  & 4.2  & 1.5  & 1.5 & 1.2  & 1.2   & 2.2  \\
\hline\hline
\end{tabular}
\end{table}

When the vector resonance $V$ is not too heavy
and if jet-tagging is eliminated, the signal
from the $W$-$V$ mixing (DY) mechanism
can have greater statistical significance
than that found above for longitudinal $W^{\pm} Z$ scattering in
the presence of jet-tagging.
Figure \ref{rhofig} shows the transverse mass distributions for
the sum of SM backgrounds and signals for
$M_V=1$ TeV and 2.5 TeV via the DY mechanism.
Despite the increase in background that results from eliminating
the jet-tag, the increase in signal event rate for a 1 TeV vector
resonance presents a clear bump near the resonance mass in
the $M_T$ spectrum.   As anticipated, the
$W^{\pm} Z$ channel  via  $W$-$V$ mixing should be the best for studying
a Vector resonance model at the LHC if $M_V \sim 1$ TeV.
Indeed, Table~\ref{lhcratesii} shows that in the
$0.85<M_T<1.05$ TeV bin
the mixing signal has a statistical significance of $S/\sqrt B\sim 15$,
far better than obtained in any of the channels after single-tagging
of the spectator jets; see Table~\ref{lhcrates}.
However, the signal rate for a 2.5 TeV vector state
is too low to be observable.
Due to the necessity of more stringent leptonic cuts to suppress
the larger SM backgrounds and the less efficient $M(\ell\ell)$ reconstruction,
the $W^+W^-$ channel seems to be less useful for observing a vector
resonance signal arising via the mixing process.
The  event rates for signals and backgrounds
are shown in Table~\ref{lhcratesii}.

A systematic comparison of the different gold-plated modes
allows one to distinguish
between the different models to a certain degree.
Models with a scalar isospin-zero resonance (SM, Scalar and O($2N$))
will yield a large excess of events in the
$ZZ\to 2\ell2\nu$, $ZZ \to 4\ell$, and $W^+W^-\to \ell^+\ell^-\nu\nu$
final states, a feature that is very distinct from predictions
of the other models; those with a vector isospin-one resonance
with $M_V \sim$ 1 TeV can be studied
most easily in the $W^\pm Z \to \ell^\pm \nu2\ell$ channel via $W$-$V$ mixing;
while models with heavier vector resonances or no resonances at all imply a
large enhancement in the $W^\pm W^\pm \to \ell^\pm \ell^\pm 2\nu$ channels.
By careful optimization of the cuts, an observable
excess of events can be seen for all of the strongly-interacting models
considered here, after several years of running of the LHC with an annual
luminosity of  $100~{\rm fb}^{-1}$.

\begin{table}[htbp]
\centering
\caption{\label{lhcratesii}%
Events per LHC-year
for $q\overline q\to W^+W^-$ and $q\overline q\to W^{\pm} Z$
channels deriving from $W$-$V$ mixing and backgrounds,
compared to the corresponding  $W_LW_L$ fusion signal rates
after removing the jet-tag  cut in Table~\protect\ref{cutstable},
assuming $\protect\sqrt s=14~{\rm TeV}$, an annual luminosity of
$100 ~{\rm fb}^{-1}$, and $m_t=175$ GeV.}
\bigskip
\begin{tabular}{l|ccc}
\hline\hline
% & Bkgd. & Vec1.0: $W$-$V$mix / fusion & Vec2.5: $W$-$V$ mix
% /  fusion\\
& Bkg & V1.0: $W$-$V$mix / fusion & V2.5: $W$-$V$ mix / fusion\\
\hline
$W^+W^-$ \hfill & 420 & 8.6 / 10 & 0.3 / 9.0 \\
%$M_{\ell\ell} > 0.5$ \hfill &  330 & 6.7 / 8.9 & 0.2 / 7.9 \cr
\hline
$W^{\pm} Z$ \hfill & 220 & 73 / 8.7 & 1.4 / 6.4 \\
\hline
$W^{\pm} Z$ \hfill & & $0.85<M_T<1.05$ TeV     &  $2<M_T<2.8$ TeV  \\
Bkg/mix/fusion &   & 22/ 69  / 3.2 &   0.82/0.81/0.55\\
\hline\hline
\end{tabular}
\end{table}

\begin{figure}[htb]
%%%\centerline{\epsfxsize=4.5in\epsfbox{mtwz3-3.eps}}
\caption{\label{rhofig}%
Transverse mass distributions for
$pp \to W^* \to V \to W^{\pm} ZX$ and $W^+ W^- X$ signals
for $M_V=1$ TeV and 2.5 TeV.  The  signal  is plotted above the
summed SM background. The mass variable of
$x$-axis is in units of GeV  and the bin size is 50 GeV.}
\end{figure}
%
%\subsubsection{Search for same-sign W-pairs at high mass ATLAS}
%

\begin{table}[htbp]
\centering
\caption{\label{TPHstrong_wpwp}%
For an integrated luminosity of 100 fb$^{-1}$,
expected rates for the \wlwl~signal and for the various background
processes and expected significances as a function of cuts.}
\bigskip
\newcommand{\lstrut}{{$\strut\atop\strut$}}
\begin{tabular}{||c||c|c|c|c||} \hline
  Process     & Same-sign & Lepton  &   Jet    & Tag      \\
              & leptons   & cuts    &  veto    & jets     \\
\hline \hline
\wlwl         &     74   &       64 &     43   &   23     \\
\wtwt         &    560   &      265 &     78   &   28     \\
g-exchange    &    380   &      175 &     14   &    1     \\
W\ttbar       &    890   &      440 &     14   &  0.3     \\
WZ/ZZ         &   1790   &      882 &    650   &   12     \\
\ttbar        &     93   &       15 &   $< 5$  & $< 5$    \\
\hline
\SB           &  0.5     &     1.5  &   1.6    & $\geq$~3.4 \\
\hline
\end{tabular}
\end{table}

The like-sign $W^+W^+$ process has been studied by the ATLAS collaboration
including detector simulations \cite{atlas,note33}.
The signal from a Standard Model Higgs boson with \mH~=~1~TeV, and the
backgrounds from \ttbar,~WZ and~ZZ were generated using
PYTHIA~5.7 \cite{RPIntro_PYTHIA}. The other background processes
were simulated at the parton level using the event generator
of Ref.~\cite{RPHstrong_Barger}.

The minimal selection cuts on the two like-sign charged leptons
are
$$
p_T(l) > 25 {\rm GeV}, \quad \quad |y(l)| < 2.5,
$$
and the event rates are shown in the first column of
Table~\ref{TPHstrong_wpwp}).
At this stage the background is overwhelming with the largest
contribution coming from~WZ/ZZ production (the \ttbar~background
has been greatly reduced by the lepton isolation cuts).
If a third lepton was present within the acceptance,
the invariant dilepton masses, computed using
all of the selected leptons of same flavour
and opposite charge, were required to be outside \mZ$\pm$15~GeV,
thus rejecting the dominant WZ/ZZ~background.
Additional cuts, which
increase the signal to background ratio, require the dilepton mass
to be above 100~GeV, that the opening angle in the transverse plane
between the two leptons be larger than 90$^\circ$,
and that their transverse momenta differ by less than 80~GeV.
The second column
of Table~\ref{TPHstrong_wpwp} shows the expected rates for the signal
and various backgrounds, after these additional lepton cuts.

To further reduce the overwhelming background,
especially the $Wt\bar t$, a jet veto is imposed for
$$
p_T(j)>40 {\rm GeV},  \quad  y(j)<2
$$
(third column of Table~\ref{TPHstrong_wpwp}).
Finally two tag jets were required in each of the forward regions,
with $15 < p_T <130$ GeV.
The upper limit set on the tag jet~\pT~significantly
reduces the \wtwt~residual backgrounds.
The last column of
Table~\ref{TPHstrong_wpwp} shows the expected rates after all
cuts. The rates for W$^-$W$^-$ pairs would be about a factor of three
lower for the \wlwl~and the \wtwt~processes, but would of course be the
same for the potentially more dangerous reducible backgrounds, thus
providing a useful control of their exact level after cuts.

   Charge misidentification is a negligible background,
given the charge identification capabilities of the
ATLAS~detector for \pT\-values between 25~and
500~GeV \cite{atlas,note33}.
In particular, it is important to note
that, after cuts, the opposite-sign lepton pairs from~W$^+$W$^-$
and \ttbar~production contain no events with leptons
of \pT~$>$~200~GeV.

\begin{figure}[htb]
% ->Removed by MG  \mbox{\epsfig{file=FPptlep.eps,width=0.95\linewidth}}
% \mbox{\epsfig{file=FPwwmlln.eps,width=0.95\linewidth}}
% put in by Han 9/6/95:
%%%\centerline{\epsfxsize=4.5in\epsfbox{ptl_atlas3-3.eps}}
\caption{\label{FPHstrong_ptlep}%
For an integrated
luminosity of 100 fb$^{-1}$, \pT -spectrum
expected for same-sign dileptons with two tag jets. Shown
are the signal expected for \mH~=~1~TeV and the various
backgrounds discussed in the text.}
\end{figure}

Figure \ref{FPHstrong_ptlep} shows the expected same-sign dilepton
\pT -distribution after all cuts. The total event rate is quite low
and the signal to background ratio does not vary much
as a function of the lepton~\pT.
Several years of
running at high luminosity will be needed to establish
an excess of events with respect to the expected background processes
in this channel.

%
%\begin{table}[htbp]
%\centering
%\caption{\label{TPHstrong_alternate}%
%Expected event rates after cuts in the \wlwl~search for an integrated
%luminosity of 100 fb$^{-1}$ and for different
%models~\protect\cite{RPHstrong_Barger}.}
%\bigskip
%\newcommand{\lstrut}{{$\strut\atop\strut$}}
%\begin{tabular}{|c|c|} \hline
%Alternate model  & Event rate \\ \hline \hline
%Standard Model (\mH~=~1~TeV)              & 23    \\  \hline
%Rescaled $\pi^+\pi^+$ scattering          & 25    \\  \hline
%Low energy theorem (LET)                  & 39    \\  \hline
%Sharp-cutoff unitarisation                & 40    \\  \hline
%$O(2N)$ Higgs-Goldstone model             & 15    \\  \hline
%\end{tabular}
%\end{table}
%

\subsection{Search for \HWWlnj~decays}

The ATLAS and CMS collaborations both studied the
very heavy Higgs searches for processes \HWWlnj~and \HZZllj~decays for
\mH~=~1000~GeV~\cite{atlas,cms}.
%\cite{atlas,RPHlnujj_note,cms}.
The study of these channels has concentrated mainly
on realistic estimates of the calorimeter performance in reconstructing
high-\pT~W/Z$\to$jj~decays in the central region (\abseta$~<$~2)
and on jet tagging in the forward regions (2~$<$~\abseta$~<$~5),
both at low and high luminosities.

   The main reason for studying the \HWWlnj~channel is that its
branching ratio is about 150~times larger than that of the gold-plated
\HZZl~channel, thus providing the largest possible signal rate
with at least one charged lepton in the final state. The \HZZllj~channel
has also been studied although it suffers from a 7~times lower rate.
The main production mechanisms for \mH~=~1000~GeV are gluon
fusion and vector boson fusion, which is dominated by \WL\WL~fusion
and contains a much smaller \ZL\ZL~contribution.
Here only vector boson fusion, qq$\to$qqH, is considered,
since the reconstruction of the two final state quark jets (or tag jets)
provides a crucial handle to extract the signal from the various
backgrounds.
%
%\subsection{Study of central region}
%
   In the reconstruction of a possible signal from heavy Higgs boson decay,
one of the most important steps is the mass reconstruction of high-\pT
W/Z$\to$jj decays, which should be as efficient and accurate as possible.
It was found \cite{atlas}
that a hadronic calorimeter granularity of at least
\detaphi~=~0.10~$\times$~0.10 is necessary to obtain
good mass resolution in the reconstruction of W$\to$~jj decays.

%The summary of
%the signal and background rates expected for an integrated
%luminosity of 100 fb$^{-1}$ and for \mH~=~1000~GeV are as follows.
%Table~\ref{TPHlnujj_HWWZZ} shows for the
%\HWWlnj~and \HZZllj~channels the production

To reconstruct the boson pair in the central region
(\abseta~$<$~2), the following acceptance cuts are imposed:

$\bullet$ \pT~$>$~100~GeV for charged leptons and neutrinos from
W/Z leptonic decays;

$\bullet$ \pT (W$\to$~l$\nu$)~$>$~350~GeV for \HWWlnj~and
          \pT (Z$\to$~ll)~$>$~200~GeV for \HZZllj;

$\bullet$ Two jets with \pT~$>$~50~GeV and with invariant mass
within $\pm$15~GeV of the W~or~Z~mass;

$\bullet$ \pT (W$\to$~jj)~$>$~350~GeV for \HWWlnj~and
          \pT (Z$\to$~jj)~$>$~200~GeV for \HZZllj.

Applying a veto on additional central
jets is expected to greatly reduce the
\ttbar~background \cite{RPHlnujj_veto,atlas}.
 A jet veto cut requiring no additional
jet with \pT~$>$~15~GeV in the central region
(\abseta~$<$~2) has an efficiency of~$\sim$~70\% for the signal
and rejects the \ttbar~(resp.~W+jet) background by a factor~$\sim$~30
(resp.~$\sim$~3). At high luminosity, the threshold on the jet veto
has to be raised to~40~GeV to maintain the same signal efficiency,
which was found to be uncertain to~$\sim~\pm$~10\%
due to uncertainties in the minimum bias model used.
The background rejection then drops to~12 (resp.~2.5) for the
\ttbar~(resp.~W+jet) backgrounds.

%
%\subsection{Study of forward region}
%

   The next step in background rejection involved the reconstruction of
tag jets~\cite{RPHlnujj_tagth} in the forward region,
i.e.~for 2~$<$~\abseta~$<$5.
Jet candidates were retained as tag jets if their
transverse energy, collected in a cone of size \dR~=~0.5, exceeds 15~GeV.
Transverse energy thresholds of 1~or~3~GeV were applied at the cell level
to study the impact of pile-up.
The forward calorimeter energy resolution has almost no impact on the
performance; the tag jet, if any, was defined as the jet with highest
energy in each of the two forward regions.
Most of the tag jets from pile-up are real jets very similar to the
low-\pT~tag jets expected from the Higgs boson signal.
No major deterioration of the tag jet reconstruction is expected
due to limitations in the integrated endcap/forward calorimeter
performance, provided the granularity is not coarser than
0.2~$\times$~0.2 in~\detaphi.

\begin{table}[htbp]
\centering
\caption{\label{TPHlnujj_tagWWZZ}%
\HWWlnj~signal and backgrounds before and after cuts in forward region
(see text). The rates are computed for an integrated luminosity of 100
fb$^{-1}$ and a lepton efficiency of 90\% has been assumed.}
\bigskip
\newcommand{\lstrut}{{$\strut\atop\strut$}}
\begin{tabular}{|c||c|c|c|c|} \hline
Process & Central & Jet  &  Single  & Double    \\
        & cuts    & veto &  tag     & tag       \\
\hline \hline
 \HWW              &   364   &  251   &   179  &  57     \\ \hline
 \ttbar            &  6520   &  560   &   110  &   5     \\
 W~+~jets          &  9540   & 3820   &   580  &  12     \\
 Pile-up           &         &        &   160  &   2     \\\hline
\SB                &  2.9    &  3.8   &   6.8  & 13.8    \\
\hline \hline
 \HZZ              &    58   &  39    &   29   &  9.1    \\ \hline
 Z~+~jets          &  1580   & 610    &  111   &  3.6    \\
 Pile-up           &         &        &   22   &  0.3    \\\hline
\SB                &   1.5   & 1.6    &  2.8   &  4.8    \\ \hline
\end{tabular}
\end{table}

\begin{figure}[htb]
%\epsfig{file=mh1000.eps,width=6cm}
%-> Fig taken out by MG\mbox{\epsfig{file=FPmh1000.eps,width=0.98\linewidth}}
% put in by Han 9/6/95:
%%%\centerline{\epsfxsize=4.5in\epsfbox{mh_atlas3-4.eps}}
\caption{\label{FPHlnujj_mh1000}%
Distribution of reconstructed Higgs boson mass for signal above
background (dashed area) for the \HWWlnj~channel and an
integrated luminosity of 100 fb$^{-1}$ (top) and for the \HZZllj~channel
and an integrated luminosity of~3$\cdot$100 fb$^{-1}$.}
\end{figure}

Table~\ref{TPHlnujj_tagWWZZ} shows,
for the \HWWlnj~and \HZZllj~signals respectively and for
an integrated luminosity of 100 fb$^{-1}$, the improvement expected in the
significance (crudely estimated as~\SB) of the signal after requiring,
first a central jet veto, and secondly
a single or double tag. The pile-up column displays separately
the total amount of background due to pile-up, but the effect
of pile-up is already included correctly in each background column.
The probability of a double tag for the signal is~$\sim$~23\%,
whereas it is only 0.62\% for the overall background, including
0.35\% of true double-tags, 0.21\% of true single-tags combined
with a single pile-up tag and 0.06\% of double pile-up tags.
The significance of the signal is obviously considerably
improved by requiring a double-tag.

   The significance of the signal improves for both channels
with the jet veto cut and mostly with the jet tag requirement.
The absence of \ttbar~background in the \HZZllj~channel does
not compensate for the much smaller rate. Figure \ref{FPHlnujj_mh1000}
shows the expected distributions for the reconstructed Higgs boson mass,
for the \HWWlnj~channel and an integrated luminosity of
100 fb$^{-1}$ (top)
and for the \HZZllj~channel and a larger integrated luminosity of
3$\cdot$100 fb$^{-1}$. The sum of the signal and background is shown
above the background (dashed area). There is not a large difference
between the shapes of the signal and background distributions,
and despite the expected significances of~13.8 for \HWWlnj~and
of~4.8 for \HZZllj, several years may be needed before a signal
can be established in this channel.

\subsection{Bounds on $L_{1,2}$ at LHC}

The authors of Ref~\cite{newherrero} have considered the bounds that can be
placed on $L_{1,2}$ by the LHC. They study the purely leptonic decays
of the $W^\pm Z$ and $ZZ$ production channels, for an integrated luminosity
of $3\times 10^5{\rm ~pb}^{-1}$. They consider all possible polarization
states of the vector bosons, so their calculation is exact in the sense that
they do not rely on the Equivalence Theorem.
On the other hand, they do not
compute the complete process at the quark parton level,
but rely on the Effective-$W$ Approximation.
They impose minimal acceptance cuts on the final state
vector bosons and use EHLQ parton distribution functions \cite{ehlq} .
They study one parameter at a time, and compare the signal with
what would be expected from the standard model with a 100~GeV Higgs boson.
They find that requiring a statistical significance
\begin{equation}
{|N(L_i)-N(M_H=100~GeV)|\over \sqrt{N(M_H=100~GeV)}} \geq 5
\end{equation}
the LHC will be sensitive to values of $L_{1,2}\sim{\cal O}(1)$.

\subsection{The Hidden Symmetry Breaking Sector at LHC}

To compute the observability of this toy symmetry breaking sector at a given
machine, the amplitude $a^{ij;kl}$ of Eq. (\ref{scatamp}) is used to derive
partonic cross sections for $W_L W_L,\ Z_LZ_L \to Z_LZ_L$ which are then
folded with the appropriate gauge boson structure functions (using the
Effective-$W$ Approximation \cite{effectw} and the EHLQ set II
\cite{ehlq} structure functions).  This yields the contribution of gauge boson
scattering to the process $pp \to ZZ + X$.

As in the standard model, gluon fusion through a top quark loop \cite{gfusion}
provides a signal for gauge boson pairs comparable to the signal from gauge
boson scattering.  The correct computation in this model is somewhat
nontrivial, since it requires the inclusion of all diagrams that contribute at
lowest nonvanishing order in $\alpha_s$ (the QCD coupling constant) and the
top quark Yukawa coupling.  To this order, there are three diagrams that
contribute to the $Z_LZ_L$ final state.  The first is the simple top quark
triangle diagram, the analogue of the Higgs boson production diagram in the
standard model.  Next there is the top-quark box, which produces final state
longitudinal $Z$'s exactly as in the standard model \cite{baur}.  Lastly,
there is a two-loop diagram, in which a box of quarks (not all four sides of
which need be top) produces a pair of Goldstone bosons which rescatter through
the Higgs boson into $Z_LZ_L$.  This last diagram must also be computed to get
a correct, gauge invariant answer, since it is leading in $1/N$.

To get a rough estimate of the gluon fusion rate, one concentrates on the
top-quark triangle, ignoring the other two diagrams.  The amplitude for this
diagram is
\be
{\alpha_s ~s~\delta^{ab} \over
{2\pi\left[ v^2 - Ns\left({1\over\lambda(M)}
+\widetilde{B}(s;m_\psi,M)\right)\right]}}
\left( g^{\mu\nu} - {2 p_2^\mu p_1^\nu \over s}\right)
I(s,m^2_t)
\punct[.]
\ee
Here $p_1$ and $p_2$ are the momenta of the two incoming gluons with
polarization vectors associated with $\mu$ and $\nu$ and colors with the $a$
and $b$ respectively, and the function $I(s, m^2_t)$ is the Feynman parameter
integral
\be
I(s,m^2_t) = m^2_t \int_0^1 dx \int_0^{1-x} dy
{(1-4 x y) \over m^2_t - xys - i\epsilon}
\punct[,]
\ee
and $m_t$ is the mass of the top quark.

\begin{figure}[htb]
% Han's change:
%%%%\centerline{\epsfxsize=4.5in\epsfbox{fig2.eps}}
%%%\centerline{\epsfbox{fig3.eps}}
\vspace{1.0in}
%\caption{\label{B}%
\caption{\label{C}%
Differential production cross section for $pp \to ZZ$ (at a
$pp$ center of mass energy of 14 TeV) as a function of invariant $Z$-pair mass
for $j=4$, $n=8$, $m_\psi = 125$ GeV and the renormalization point $M=2500$
GeV. A rapidity cut of $|y| < 2.5$ has been imposed on the final state
$Z$s. The gauge boson scattering signal is shown as the dot-dash curve and
gluon fusion signal (with $M = 4300$ GeV) as the solid curve.  The background
from $q \bar q$ annihilation is shown as the dashed curve.  In all
contributions, the rapidities of the $Z$s must satisfy $|y_Z|<2.5$.  All
computations use the EHLQ set II structure functions with $Q^2 = M_W^2$ in the
gauge boson scattering curve, and $Q^2 = \hat s$ in the other two cases.}
\end{figure}

In Figure \ref{C} is shown the differential cross section for $ZZ$ production
at the LHC as a function of $ZZ$ invariant mass.  The $Z$ bosons both must
have rapidity less than 2.5.  Here $n=8$ and $M=2500$ GeV
(with a Higgs boson mass of approximately 485 GeV).
The gauge boson
fusion contribution is shown in the dot-dash curve, and the gluon fusion
contribution is shown as the solid curve.
The irreducible background to observing the Higgs boson in $Z$ pairs comes
from the process $q \bar{q} \to ZZ$ and is shown as the dashed curves.
The gauge boson scattering signal here is ``hidden''
because the Higgs boson is both light and broad.  The gluon
fusion signal is substantially larger than the gauge boson scattering
signal. The signal, however, is still significantly below the background,
making detection of a broad resonance difficult.

\begin{figure}%[htb]
%%%\centerline{\epsfbox{fig5.eps}}
\vspace{0.5in}
\caption{\label{E}
Same as the previous figure, but
for a standard model Higgs boson with mass 485 GeV.}
\end{figure}

By way of comparison, the signal for a 485 GeV standard model Higgs boson is
shown in Figure \ref{E}.
In this case, because the Higgs boson is relatively narrow,
{\it on the peak} the gauge boson scattering Higgs boson signal is comparable
to the background and the gluon fusion signal is well above the background.

In \cite{ny,kny}\ it was shown that the numbers of final state gauge boson
pairs from gauge boson scattering is roughly independent of $N$ if $\sqrt{N}
M$ is held fixed. This is because as $N$ increases for fixed $\sqrt{N} M$, $M$
and the mass and width of the Higgs boson decrease like $1/\sqrt{N}$. The
increased production of Higgs bosons due to their smaller mass\footnote{And
therefore higher gauge boson partonic luminosity \cite{changail,effectw}.} is
approximately cancelled by the Higgs boson's smaller branching ratio into $W$s
and $Z$s. The number of signal events, therefore, is approximately independent
of $N$ and is the same as the number which would be present in the model with
$n=0$. Since the signal for gauge boson scattering in that model is (perhaps)
observable \cite{changail} and since the number of $ZZ$ events is roughly
independent of $n$, the authors of \cite{ny,kny} argue that the signal may
be observable for any $n$\footnote{Note though that the signal to background
ratio is not as good at the 14 TeV LHC as it would have been at the 40 TeV
SSC \cite{hpenom}.}.

%\begin{figure}[htb]
%\centerline{\epsfbox{fig3.eps}}
%\vspace{0.5in}
%\caption{\label{C}
%Same as the previous figure, but with $n=8$ and $M=2500$ GeV}
%\end{figure}
%
%\begin{figure}[htb]
%\centerline{\epsfbox{fig4.eps}}
%\vspace{0.5in}
%\caption{\label{D}
%Same as the previous figure, but with $n=0$ and $M=4300$ GeV.}
%\end{figure}

%There are two technical shortcomings in the calculation of the cross sections
%presented here.  Firstly, in computing the gauge boson scattering signal, one
%uses both the Equivalence Theorem and the Effective-$W$
%Approximation \cite{effectw}.  Strictly speaking, both of these
%approximations hold only at energies above a few times the $W$ mass. Even at
%200 GeV, however, the corrections should be of order one \cite{gunion}.
%However, the background at these energies when $n = 32$ is more than an order
%of magnitude larger than the signal.  While one cannot precisely determine the
%number of gauge boson scattering events, it is clear that they are swamped by
%the background.  Second, as discussed above, this computation of gluon fusion
%included only the contribution from a top quark triangle diagram and not the
%contributions from the other two of the three leading diagrams.  All three
%diagrams are necessary, because they interfere.  However, it is reasonable to
%expect that their proper inclusion will not change any of the conclusions.
%

In general, the two-gauge-boson scattering signal of a symmetry breaking
sector is not visible above the background unless the gauge boson elastic
scattering amplitudes are big. This happens either when the symmetry breaking
sector is strongly coupled or at the peak of a narrow resonance.  It is
possible to see this just by counting coupling constants: the background ($qq
\to ZZ$) is order $g^2$ and is a two body final state, while the
signal ($qq \to qqZZ$) is naively of order $g^4$ and is a four body
final state. When the symmetry breaking sector is strongly interacting and the
final state gauge bosons are longitudinal, this naive $g^4$ gets replaced by
$g^2 a^{ij;kl}$, and the signal may compete with the background. Since
$a^{ij;kl}$ in the $O(4+32)$ model is {\it never} large, the signal rate {\it
never approaches} the background rate.

Moreover, while this section is focussed on the signal for the $ZZ$ final
state, the arguments given here should apply equally to all other
two-gauge-boson signals as well. In the $O(4+32)$ model it is likely that {\it
none} of the two-gauge-boson signals of the symmetry breaking sector may be
observed over the background.  In this model even observing all
two-gauge-boson modes will not be sufficient to detect the dynamics of
electroweak symmetry breaking -- one will need to observe the pseudo-Goldstone
bosons, and identify them with symmetry breaking.

\clearpage

\section{Strongly-Interacting ESB Sector at $e^+e^-$
%\hfill\break
Linear Colliders}

\subsection{Vector Resonance Signals at LEP-II and NLC}

Future $e^+e^-$ colliders are sensitive to the neutral
vector resonance  if the mass
 $M_V$ of the new boson multiplet lies not
far from the maximum machine energy; or if it is lower,
such a resonant contribution would be quite manifest.
If the masses  of the $V$ bosons are  higher than the maximum c.m.
energy,  they give rise to
indirect effects in the
$e^+e^-\to f^+f^-$ and  $e^+e^-\to W^+W^-$  cross sections.
The result of the analysis in Ref.~\cite{lin}
is that virtual effects of the vector state are also important.
It appears that annihilation into a fermion pair in such  machines, at the
considered luminosities, would marginally improve on existing limits
if polarized beams are available and left-right asymmetries are measured.
On the other hand, the process of $W$-pair production in the
$e^+e^-$ annihilation would allow for sensitive tests of the
strong sector,
especially if the $W$ polarizations are  reconstructed from their
decay distributions.
% and the more so the higher the energy of the machine.
%This is because BESS modifies the standard couplings  in such
%a way that the typical cancellation present in the SM do not happen
%anymore and the amplitude is growing with $s$.

Ref.~\cite{lin}  analyzes
cross-sections and asymmetries for the channel $e^+e^-\to f^+f^-$
and $e^+e^-\to W^+W^-$, assuming that it will
be possible to separate
$e^+e^-\to W^+_L W^-_L$, $e^+e^-\to W^+_L W^-_T$, and
$e^+e^-\to W^+_T W^-_T$. The distribution of the $W$ decay angle
in the c.m. frame depends indeed in a very distinct way on its helicity,
being peaked forward (backward) with respect to the production direction
for positive (negative) helicity or at $90^o$ for zero helicity.

Consider now the $WW$ channel, for one $W$  decaying leptonically
and the other hadronically.
To discuss the restrictions on the parameter space for masses of the
resonance a little higher than the available energy, the calculation takes into
account the experimental efficiency.  It is assumed that there will be
an overall detection efficiency of 10\% including
the branching ratio
$B=0.29$ and the loss of luminosity from beamstrahlung.

For a collider at $\rs=500$ GeV with an integrated luminosity
of 20 fb$^{-1}$ the results are illustrated in Figure \ref{allowed}.
The contours have
been obtained by taking 18 bins in the angular region restricted by
$|\cos\theta|< 0.95$. This figure illustrates the 90\% C.L. allowed regions
for $M_V=600$ GeV
obtained by considering the unpolarized $WW$ differential cross-section
(dotted line), the $W_LW_L$ cross section (dashed line),
and the combination of the left-right asymmetry with all the
differential cross-sections for the different final $W$ polarizations
(solid line). It is clear that even
at the level of the
unpolarized cross-section there are important improvements
with respect to LEP-I.
\begin{figure}[htb]
%
%\epsffile[69 263 498 700]{contr_bess4-1.ps}
%%%\centerline{\epsfxsize=8truecm\epsfbox{contr_bess4-1.eps}}
\caption{\label{allowed}
90\% C.L. allowed regions for $M_V=600~GeV$
obtained by considering the unpolarized $WW$ differential cross-section
(dotted line), the $W_LW_L$ cross section (dashed line),
and the combination of the left-right asymmetry with all the
differential cross-sections for the different final $W$ polarizations
(solid line).}
\end{figure}

For colliders with $\rs=1,~2$ TeV
and for $M_V=$1.2 and 2.5 TeV respectively,
the allowed region, combining all the observables,
 reduces in practice to a line.
Therefore, even the unpolarized $WW$ differential
cross section measurements
can improve the bounds.

%{\it Barklow's}

The effects of a vector resonance can be incorporated also by
multiplying the standard model amplitude for $e^+e^-\to\wwl$
by the complex form factor $F_T$ \cite{mpeskin,barklow}
where\footnote{This discussion was provided by T. Barklow.}
\begin{eqnarray}
 F_T &=&
         \exp\bigl[{1\over \pi} \int_0^\infty
          ds'\delta(s',M_\rho,\Gamma_\rho)
          \{ {1\over s'-s-i\epsilon}-{1\over s'}\}
         \bigr],  \\
\delta(s) &=& {1\over 96\pi} {s\over v^2}
+ {3\pi\over 8} \left[ \tanh (
{
s-M_\rho^2
\over
M_\rho\Gamma_\rho
}
)+1\right],
\end{eqnarray}
%
%$v=240$~GeV,
$M_\rho$ is the
techni-$\rho$ mass and $\Gamma_\rho$ is the techni-$\rho$ width.
Note that for an infinite techni-$\rho$ mass $\delta(s)$ becomes
\begin{equation}
\delta(s)= {1\over 96\pi} {s\over v^2},
\end{equation}
reflecting the low energy theorem (LET) amplitude for longitudinal
gauge boson scattering.

%
% put in by Han 9/6/95:
%\epsffile[188 275 410 501]{contr_bklw4-1.eps}
%
\begin{figure}[htb]
%%%\centerline{\epsfxsize=4.5in\epsfbox{contr_bklw4-1.eps}}
%\centerline{\epsfxsize=4.5in\epsfbox{bklw_scan.eps}}
\caption{Confidence level contours for the real and imaginary parts of
$F_T$ at $\protect \sqrt{s} = 1500$~GeV with 190 fb$^{-1}$.
The initial state electron polarization is 90\%. The contour about the
light Higgs boson value of $F_T=(1,0)$ is 95\% confidence level and
the contour about the $M_\rho=4$~TeV point is 68\% confidence level.}
\label{contours}
\end{figure}

Figure \ref{contours} contains confidence level contours for the real and
imaginary parts of $F_T$ at $\sqrt{s}=1500$ GeV with 190
fb$^{-1}$ \cite{barklow}.
Shown are the 95\% confidence level contour about the
light Higgs boson value of $F_T$, as well as the 68\% confidence level
(i.e., $1\sigma$ probability) contour about the value of $F_T$
for a 4 TeV techni-$\rho$.
Even the non-resonant LET point is well outside the light Higgs boson
95\% confidence level region.
%In fact, the LET point intersects the
%99.99945\% confidence level contour about the light Higgs boson point,
%corresponding to a 4.5$\sigma$ signal.
The  6~TeV and
and 4~TeV techni-$\rho$ points correspond to 4.8$\sigma$ and
6.5$\sigma$ signals, respectively.
At a slightly higher integrated luminosity
of 225 fb$^{-1}$, it is possible to obtain 7.1$\sigma$, 5.3$\sigma$ and
5.0$\sigma$ signals for
a 4~TeV techni-rho, a 6~TeV techni-rho, and LET, respectively.

%Furthermore, it is found \cite{barklow} that the chiral
%Lagrangian parameters $L_{9L}$ and $L_{9R}$ can also be determined
%with an accuracy of $\pm 1.5$ at $\sqrt{s}=500$~GeV with 80 fb$^{-1}$,
%and $\pm 0.5$ at $\sqrt{s}=1500$~GeV with 190 fb$^{-1}$ (95\% C.L.).

\subsection{$WW$ Fusion Processes: \hfill\break
$W^+ W^- \to W^+W^-$  versus $W^+ W^- \to ZZ$}

%{\it Dijet mass resolution}

The ratio of $W^+ W^- \to W^+ W^-$ and $W^+W^- \to ZZ$
cross sections is a sensitive probe of the strongly-interacting
electroweak sector \cite{HanHawaii,epemww}, since
the models have distinctive particle spectra with different
weak isospin content.  For a scalar-dominance model,
one expects the $W^+_L W^-_L$ rate to be larger than
$Z_LZ_L$,  {\it  e.g.}  a SM-like Higgs boson dominating
in the $s$-channel gives
$\sigma(H \rightarrow W^+_LW^-_L)/\sigma(H \rightarrow  Z_LZ_L) \sim$ 2.
For a vector-dominance model there would be a significant
resonant enhancement  in the $W^+_L W^-_L$ mode, but not in $Z_L Z_L$
due to the weak isospin conservation in  strongly-interacting
electroweak sector
(just like $\rho^0 \to \pi^+ \pi^-$ but not $\pi^0\pi^0$ in QCD).
On the other hand, if the resonances
are far from our reach, then the LET amplitudes behave like
$-u/v^2$ for $W_L^+W_L^-\to W_L^+W_L^-$
and like $s/v^2$ for $W_L^+W_L^-\to Z_LZ_L$, so that
$\sigma(W_L^+W_L^- \to Z_LZ_L)/\sigma(W_L^+W_L^- \to W_L^+W_L^-)=3/2$.
The $Z_L Z_L$ rate is then larger than $W^+_LW^-_L$,
and even more so in the central scattering region.
Measuring the relative yields of $W^+_L W^-_L$ and $Z_L Z_L$
will therefore reveal important characteristics of the
strongly-interacting ESB sector.

The $W^\pm$ and $Z$ bosons may be detected by their dijet
decay modes and identified via the dijet invariant masses
$M(W^\pm \to jj)\simeq M_W$, $M(Z\to jj)\simeq M_Z$.  With realistic mass
resolution, discrimination cannot be made event-by-event but
can be achieved on a statistical basis.

   The experimental $W$ dijet mass distributions
will contain the intrinsic decay widths folded with experimental
resolution factors depending on calorimetry and geometry.
It is possible to explore \cite{epemww}
the dijet mass resolution using two alternative jet
energy resolution algorithms~\cite{jlc}
\begin{eqnarray}
\delta E_j/E_j &=& 0.50 \Big/ \sqrt{E_j} \;\oplus\; 0.02 \qquad
\mbox{Algorithm A} \\
               &=& 0.25 \Big/ \sqrt{E_j} \;\oplus\; 0.02 \qquad
\mbox{Algorithm B}
\end{eqnarray}
\begin{figure}[htb]
% put in by Han 9/6/95:
\centerline{\epsfxsize=8truecm
\epsffile[69 263 498 700]{mjj_4-2.ps}}
\caption{\label{dijetmass}%
$W^\pm \to jj$ and $Z\to jj$ dijet invariant mass distributions for
$e^+e^- \to e\nu WZ$ events at $\protect\sqrt s=1.5$ TeV,
found by applying
(a)~algorithm~A and (b)~algorithm~B (see text) for calorimeter energy
resolution, omitting angular resolution and heavy-quark decay  effects.
}
\end{figure}

\noindent
in GeV units, where the symbol $\oplus$ means adding in quadrature.
This was applied
to the typical SM background process $e^+e^-\to e^+\nu W^-Z$ at $\sqrt s =
1.5$~TeV, averaging over all final $W\to jj$
%and $Z\to jj$
dijet decays with
%the acceptance cuts of Section IV and
gaussian smearing of jet energies according to these algorithms; the resulting
$W^\pm \to jj$ and $Z\to jj$ dijet invariant mass distributions are shown in
Figure \ref{dijetmass}. Since this study omits angular resolution effects,
sensitive to details of detector design, in the illustrations below the
more conservative algorithm~A is adopted.

 If one now identifies dijets having measured mass in the intervals
$$[0.85M_W, \; {1\over 2}(M_W+M_Z)] \quad
{\rm and}  \quad [{1\over 2}(M_W+M_Z), \; 1.15M_Z]$$
as $W^\pm \to jj$ and $Z\to jj$, respectively, and  includes the effects of
semileptonic decays of $b$ and $c$ quarks, algorithm A indicates that
true $W^+ W^-$, $W^\pm Z$, $ZZ\to jjjj$ events will be interpreted
statistically as follows:
$$
\begin{array}{lcrrrrr}
WW &\Rightarrow & 73\%\: WW, & 17\%\: WZ, &  1\%\: ZZ,&  9\%\: {\rm reject},
\\
WZ &\Rightarrow & 19\%\: WW, & 66\%\: WZ, & 7\%\: ZZ,&  8\%\: {\rm reject},
\\
ZZ &\Rightarrow &  5\%\: WW, & 32\%\: WZ, & 55\%\: ZZ,&  8\%\: {\rm reject},
\end{array}
$$
These numbers show that misidentification of $W^+ W^-$
as $ZZ$ (or vice versa)
is very unlikely; also the loss of $W^+ W^-$ or $ZZ$ signal strength is not
in itself very serious. The principal danger comes from $W^\pm Z$ events
that are misidentified as $W^+ W^-$ or $ZZ$, confusing or even swamping
these signals if $W^\pm Z$ production is relatively large.  One must therefore
ensure, via suitable acceptance criteria, that $W^\pm Z$ production is not
an order of magnitude bigger than $W^+ W^-$ or $ZZ$ signal.

%
%\section{SM calculations and acceptance cuts}
%

The SM signals for  $W_L^+ W_L^-\to W_L^+W_L^-,\; Z_LZ_L$ fusion processes
with a heavy Higgs boson have been considered previously, along with certain
SM backgrounds~\cite{fusionjfg,hagi,fusionvb,najima}.
The irreducible SM backgrounds to the Strongly-interacting
Electroweak Sector,  which include transversely polarized
vector bosons  $W_T^\pm$ and  $Z_T$ production,
can be obtained by setting $m_H = 0$; further backgrounds
arise from misidentifying other $W$ and $Z$ combinations.

\begin{figure}[htb]
% put in by Han 9/6/95:
\centerline{\epsfxsize=8truecm
\epsffile[69 263 498 700]{sigma_4-2.ps}}
\caption{\label{fig:sews}%
Cross sections for SM scattering processes that can contribute
to strongly-interacting EWSB
signals and backgrounds in the $e^+e^-\to \bar\nu\nu W^+W^-$ and
$\bar\nu\nu ZZ$ channels, versus CM energy $\protect\sqrt s$.
}
\end{figure}

The remaining scattering cross sections of interest are illustrated
in Figure \ref{fig:sews}.  This figure shows SM cross sections for the fusion
processes with both $m_H=0$ (solid curves) and $m_H=1$ TeV (dashed
curves);  the excess over the $m_H=0$
case represents the  strongly-interacting EWSB
signal in the SM with a heavy Higgs boson.
The  signals of present interest have final-state $W_L^+ W_L^-$ and
$Z_LZ_L$ pairs, giving four-jet final states with two undetected
neutrinos.

\begin{figure}%[htb]
%put in by Han 9/6/95:
%%%\centerline{\epsfxsize=8truecm\epsfbox{events_4-2.eps}}
%\centerline{\epsfxsize=4.5in\epsfbox{events_4-2.eps}}
%\epsffile[69 263 498 700]{events_4-2.ps}}
\caption{\label{fig:events}%
Expected numbers of $W^+W^-, ZZ \to (jj)(jj)$ signal and background
events, in 20 GeV bins of diboson invariant mass, for 200~fb$^{-1}$
luminosity at $\protect\sqrt s=1.5$ TeV:
(a) $W^+W^-$ events, (b) $ZZ$ events.
Dijet branching fractions and $W^\pm/Z$ identification/misidentification
factors are included.  The dotted histogram denotes total SM
background including misidentifications.  The solid, dashed and dot-dashed
histograms denote signal plus background for the LET, SM and CCV models,
respectively; CCS model results are close to the SM~case.
}
\end{figure}

The basic acceptance cuts are the following.  Since one is interested
in $WW$ scattering at high subprocess energy, one looks for
pairs of weak bosons with high invariant masses $M_{WW}$, high transverse
momenta $p_T(W)$ of the vector bosons,
and relatively large angles $\theta_W$ with respect to the beam axis.
The cuts require

\begin{equation}
M_{WW} >500{\rm\ GeV} \;;
\quad p_T(W) > 150{\rm\ GeV} \;;
\quad |\cos\theta_W| <0.8.
\label{eq:level1}
\end{equation}

The SM $e^+e^-W^+W^-$ background gets very large contributions from
the virtual $\gamma\gamma\to W^+W^-$ subprocess, which gives
mainly dibosons with small net transverse momentum $p_T(WW)$, quite
unlike the $W_LW_L$ signal and other backgrounds.
It is demonstrated \cite{epemww} that it is advantageous to select
an intermediate range of $p_T(WW)$, to remove a lot of background
at little cost to the signal;  somewhat similar cuts are made
for $p_T(ZZ)$ , though these are less crucial.  Specifically, the cuts
require
\begin{equation}
50{\rm\ GeV} < p_T(WW) < 300{\rm\ GeV}, \; \; \;
20{\rm\ GeV} < p_T(ZZ) < 300{\rm\ GeV},
\label{eq:level2}
\end{equation}
at $\sqrt s =1.5$ TeV.
With large minimum $p_T(WW)$ and $p_T(ZZ)$ requirements, it becomes
much less likely that the final-state electrons in $eeWW$ and $e\nu WZ$
background channels can escape undetected down the beam-pipes; a veto
on visible hard electrons is now very effective against $eeWW$ (less so
against $e\nu WZ$).  Therefore, the veto \cite{hagi}
\begin{equation}
\mbox{no $e^\pm$ with $E_e>50$ GeV and} \quad
|\cos\theta_e| < \cos(0.15\rm\ rad)
\label{eq:level3}
\end{equation}
is imposed.

\begin{table}[htbp]
\centering
\def\arraystretch{.8} %% Necessary to keep table from going off the page
\caption{\label{tablei}%
Cross sections in fb, before and after cuts, for $e^+e^-$ collisions
at $\protect\sqrt s=1.5$ TeV.
For comparison, results for $e^-e^- \to \nu \nu W^-W^-$
are also presented, with the same energy and the $W^+W^-$ cuts.
Hadronic branching fractions of $WW$ decays and the
$W^\pm/Z$ identification/misidentification  are not included here.
The first number in the final $e^+e^-W^+W^-$ and $e\nu WZ$ entries
denotes the $p_T > 20$~GeV choice, for the case where $WW$ and $WZ$
are misidentified as $ZZ$; the second number (in parentheses) denotes
the $p_T > 50$~GeV choice, for the case where they are identified as $WW$.}
\[
\begin{array}{|l|c|c|c|}
\hline
\mbox{Contribution}&\mbox{no cuts}&
\mbox{$+$ Eq.~(\ref{eq:level1})}&
\mbox{$+$ Eqs.~(\ref{eq:level1})--(\ref{eq:level3})}\\
\hline
\bar\nu\nu W^+W^-\mbox{ signals (fb)} & & & \\
\quad\mbox{SM }(m_H^{}=1\mbox{ TeV})& 7.7   & 3.5      & 2.4 \\
\quad\mbox{CCS }(M_S^{},\Gamma_S=1,0.35 \mbox{ TeV})  & - & 3.5  &2.4  \\
\quad\mbox{CCV }(M_V^{},\Gamma_V=1,0.03 \mbox{ TeV})  & - & 1.5   & 1.0  \\
\quad\mbox{LET }(m_H^{}=\infty)&3.1 & 0.61         & 0.46              \\
\hline
\bar\nu\nu ZZ\mbox{ signals (fb)} & & & \\
\quad\mbox{SM }(m_H^{}=1\mbox{ TeV})& 5.9   & 2.4  & 2.2 \\
\quad\mbox{CCS }(M_S^{},\Gamma_S=1,0.35 \mbox{ TeV})  & - & 2.7  &   2.5 \\
\quad\mbox{CCV }(M_V^{},\Gamma_V=1,0.03 \mbox{ TeV})  & - &  0.72  & 0.67 \\
\quad\mbox{LET }(m_H^{}=\infty)&3.4 & 0.89          & 0.84            \\
\hline
\nu\nu W^-W^-\mbox{ signals (fb)} & & & \\
\quad\mbox{SM }(m_H^{}=1\mbox{ TeV})& 2.7  & 0.53      & 0.39 \\
\quad\mbox{CCS }(M_S^{},\Gamma_S=1,0.35 \mbox{ TeV})  & - & 0.71    & 0.52   \\
\quad\mbox{CCV }(M_V^{},\Gamma_V=1,0.03 \mbox{ TeV})  & - & 0.72 & 0.53 \\
\quad\mbox{LET }(m_H^{}=\infty)& 3.5 &   0.89 & 0.63  \\
\hline
\mbox{SM Backgrounds (fb)} & & & \\
\quad\bar\nu\nu W^+W^-\;(m_H^{}=0)    & 45 & 1.1 & 0.86 \\
\quad\bar\nu\nu ZZ\;(m_H^{}=0)    & 18  & 0.84 & 0.72 \\
\quad e^+e^-W^+W^-\;(m_H^{}=0)& 2000  & 28 & 3.5 (0.95) \\
\quad e\nu WZ\;(m_H^{}=0)     &  150  & 4.6  & 3.1 (2.7) \\
\quad e^-e^- \to \nu \nu W^-W^-  \;
(m_H^{}=0)& 51  & 2.3 & 1.7 \\
\hline
\end{array}
\]
\end{table}

Table~\ref{tablei} presents the results  for $e^+e^-$ collisions at $\sqrt s =
1.5$ TeV, showing signal and  background cross sections before and
after successive cuts.   Here the SM with a heavy Higgs boson and LET Model
signals have been found by subtracting the SM $m_H=0$ intrinsic background
from SM $m_H=1$ TeV and $m_H=\infty$ values, respectively. Partial wave
unitarity is respected at all energies reached here
so that no unitarization needs to be imposed.
%~\footnote{
%We have employed the simple Breit-Wigner form for the
%Higgs boson propagator with its tree-level width. This prescription, however,
%may violate unitarity at the heavy Higgs boson resonance and the
%cross section
%may be overestimated by about 20\%. See \protect{\cite{peak}}.}
For the chirally coupled scalar (CCS) and chirally coupled
vector (CCV) models, the signals are calculated in the Effective $W$-boson
Approximation.
The validity of this approximation can be checked by
comparing CCS (with $g_S=1$) to the  exact SM  results;
there is agreement at the $20\%$ level,  using the
cuts in Eq.~(\ref{eq:level1}).  In such an approximation, however, the
kinematical cuts
of Eqs.~(\ref{eq:level2})--(\ref{eq:level3}) cannot be implemented;
it is
therefore assumed that
the efficiencies of these cuts are the same as for the SM with a heavy Higgs
boson ($m_H=1$ TeV) signal.
For comparison, results for  $e^-e^- \to \nu\nu W^-W^-$
are also included \cite{bbch}, with the same cuts as the $\bar\nu\nu W^+W^-$
case.   Note that the LET signal rates for  $e^+e^-\to \nu\bar\nu ZZ$ and
$e^-e^-\to \nu\nu W^-W^-$ channels are essentially equal
(when the cuts imposed are
the same); this is a consequence of  the Low Energy Theorem and crossing
symmetry for $W_LW_L$ scattering.
Branching fractions  for $W \to jj$ decays and
$W^\pm/Z$  identification/misidentification factors
are not included in this table.

Figure \ref{fig:events} shows the expected signal
and background event rates versus diboson mass for different models at
a 1.5~TeV NLC, assuming an integrated luminosity of 200 fb$^{-1}$.
The branching fractions  $BR(W \to jj)=67.8\%$ and $BR(Z\to jj)=69.9\%$
\cite{pdg} and the
$W^\pm /Z$  identification/misidentification factors
%(final set of Section~III)
are all included here.
Comparing the $W^+ W^-$ events [Fig.~\ref{fig:events}(a)]
and $ZZ$ events [Fig.~\ref{fig:events}(b)],
it is again clear that a broad Higgs-like scalar will enhance both
$W^+ W^-$
and $ZZ$ channels with $\sigma(W^+ W^-) > \sigma(ZZ)$; a $\rho$-like vector
resonance will manifest itself through $W^+W^-$ but not $ZZ$; while the LET
amplitude will enhance $ZZ$ more than $W^+ W^-$.
Table~\ref{tableii} summarizes the corresponding total signal $S$ and
background $B$
event numbers, summing over diboson invariant mass bins,  together with
the statistical significance $S/\sqrt B$.  The  LET signal for  $W^+W^-$  is
particularly small; the ratio $S/B$ can be enhanced by making a higher
mass cut (e.g.\ $M_{WW} > 0.7$ TeV), but the significance $S/\sqrt B$
is not in fact improved by this.
Results for  $e^-e^- \to \nu\nu W^-W^-$ have again been
included for comparison.

At the NLC, since electron polarization of order 90--95\% at injection
with only a few percent depolarization during acceleration may
well be achievable \cite{nlcetc},
it is interesting to consider also the effects of beam polarization.
The $W^+W^- \to W^+ W^-,ZZ$ scattering signals of interest arise
from initial $e^-_L$ and $e^+_R$ states only and the signal cross sections
are therefore doubled with an $e^-_L$ beam.
Table~\ref{tableiii}(a)  shows the background
cross sections for the beams $e^+ e^-_L$, $e^- e^-_L$
and $e^-_L e^-_L$.  Based on these results, event numbers and significances
for the case of 100\% $e^-_L$ beam at  $\sqrt s=1.5$ TeV with 200 fb$^{-1}$
are shown in Table~\ref{tableiii}(b), to be compared with Table~\ref{tableii};
$S$ and $B$ for intermediate beam polarizations
can be found by interpolating Table~\ref{tableii} and Table~\ref{tableiii}.

\begin{table}[htbp]
\centering
\caption{\label{tableii}%
Total numbers of $W^+W^-, ZZ \to  4$-jet
signal $S$ and background $B$ events calculated for  a 1.5~TeV
NLC with  integrated luminosity 200~fb$^{-1}$.  Events are summed
over the mass range $0.5 < M_{WW} < 1.5$~TeV except for the $W^+W^-$ channel
with  a narrow vector resonance in which $0.9 < M_{WW} < 1.1$~TeV. The
statistical significance $S/\protect\sqrt B$ is also given.
For comparison, results for $e^-e^- \to \nu \nu W^-W^-$
are also presented, for the same energy and luminosity and the $W^+W^-$
cuts. The hadronic branching fractions of $WW$ decays and the $W^\pm/Z$
identification/misidentification are included.}
\bigskip
\begin{tabular}{|l|c|c|c|c|c|} \hline
channels & SM  & Scalar & Vector   & LET  \\
\noalign{\vskip-1ex}
& $m_H=1$ TeV & $M_S=1$ TeV & $M_V=1$ TeV &\\
\hline
$S(e^+ e^- \to \bar \nu \nu W^+ W^-)$
& 160   & 160   & 46  & 31  \\
$B$(backgrounds)
& 170    & 170   & 4.5  & 170  \\
$S/\sqrt B$ & 12 & 12 & 22 & 2.4 \\
\hline
$S(e^+ e^- \to \bar\nu \nu ZZ)$
&  120  & 130  & 36  & 45   \\
$B$(backgrounds)
& 63    & 63   & 63  & 63  \\
$S/\sqrt B$ & 15& 17& 4.5& 5.7\\
\hline
\hline
$S(e^- e^- \to \nu \nu W^- W^-)$
& 27  & 35  & 36  & 42  \\
$B$(backgrounds)
& 230  & 230   & 230  & 230  \\
$S/\sqrt B$ & 1.8 & 2.3 & 2.4 & 2.8 \\
\hline
\end{tabular}
\end{table}

\begin{table}[htbp]
\def\arraystretch{.8}
\centering
\bigskip
\caption{\label{tableiii}%
Improvements from using $100\%$ polarized $e^-_L$ beams in a 1.5~TeV
$e^+e^-/e^-e^-$ collider.
Part (a) gives SM background cross sections in fb with the full cuts
Eqs.~(\protect{\ref{eq:level1}})--(\protect{\ref{eq:level3}}); the signal
cross sections are
simply doubled with each $e^-_L$ beam  compared to
Table~\protect{\ref{tablei}}.
Part (b) gives the expected numbers of signal and background events
for integrated luminosity 200~fb$^{-1}$, to be compared with
Table~\protect{\ref{tableii}}.
}
\bigskip
\begin{tabular}{|l|c|}
\hline
(a) SM Backgrounds & Cross sections in fb with
Eqs.~(\ref{eq:level1})--(\ref{eq:level3}) \\
\hline
$\quad e^+e^-_L \to \bar \nu \nu W^+ W^- \; (m_H=0)$ & 1.7 \\
$\quad e^+e^-_L \to \bar \nu\nu ZZ \; (m_H=0)$   &    1.4 \\
$\quad e^+e^-_L \to e^+e^- W^+W^-\; (m_H=0)$ &         4.3 (1.3) \\
$\quad e^+e^-_L \to e\nu WZ\; (m_H=0)$ &               4.5 (3.9) \\
\hline
\hline
$\quad e^-e^-_L \to \nu \nu W^- W^- \; (m_H=0)$ &  3.4 \\
$\quad e^-e^-_L \to e^-e^- W^+ W^- \; (m_H=0)$ &  1.3 \\
$\quad e^-e^-_L \to e^-\nu  W^- Z \; (m_H=0)$ &   4.4 \\
\hline
$\quad e^-_L e^-_L \to \nu \nu W^- W^- \; (m_H=0)$ &  6.8 \\
$\quad e^-_L e^-_L \to e^-e^- W^+ W^- \; (m_H=0)$ &  1.8 \\
$\quad e^-_L e^-_L \to e^-\nu  W^- Z \; (m_H=0)$ &   6.5 \\
\hline
\end{tabular}

\vspace{0.05in}

\begin{tabular}{|l|c|c|c|c|}
\hline
(b) channels & SM  & Scalar & Vector   & LET  \\
& $m_H=1$ TeV & $M_S=1$ TeV & $M_V=1$ TeV &\\
\hline
$S(e^+ e^- \to \bar \nu \nu W^+ W^-)$
& 330   & 320   & 92  & 62  \\
$B$(backgrounds)
& 280    & 280   & 7.1  & 280  \\
$S/\sqrt B$ & 20 & 20 & 35 & 3.7 \\
\hline
$S(e^+ e^- \to \bar\nu \nu ZZ)$
&  240  & 260  & 72  & 90   \\
$B$(backgrounds)
& 110    & 110   & 110  & 110  \\
$S/\sqrt B$ & 23 & 25& 6.8& 8.5\\
\hline
\hline
$S(e^- e^-_L \to \nu \nu W^- W^-)$  & 54 & 70 & 72 & 84 \\
$B$(background) & 400 & 400 & 400 & 400\\
$S/\sqrt B$ & 2.7 & 3.5 & 3.6 & 4.2 \\
\hline
$S(e^-_L e^-_L \to \nu \nu W^- W^-)$  & 110 & 140 & 140 & 170 \\
$B$(background) & 710 & 710 & 710 & 710\\
$S/\sqrt B$ & 4.0 & 5.2 & 5.4 & 6.3 \\
\hline
\end{tabular}
\end{table}

Since the strongly-interacting EWSB signals increase with  CM energy,
a  2 TeV $e^+e^-$ linear collider  would give a larger signal rate.
At $\sqrt s=2$ TeV the $m_H=1$ TeV and $m_H=\infty$
signal cross sections are, for $W^+W^-$, (see Figure \ref{fig:sews})
\begin{eqnarray}
\sigma^{}_{\rm SEWS}({\rm SM}\; 1 {\rm TeV}) &=& \sigma_{W^+W^-}
(m_H=1 \; {\rm TeV})
- \sigma_{W^+W^-}(m_H=0) \simeq 20 \; {\rm fb} \nonumber \\
\sigma^{}_{\rm SEWS}({\rm LET}) &=& \sigma_{W^+W^-}(m_H=  {\infty})
- \sigma_{W^+W^-}(m_H=0) \simeq  5 \; {\rm fb}  \nonumber
\end{eqnarray}
and for $ZZ$,
\begin{eqnarray}
\sigma^{}_{\rm SEWS}({\rm SM}\; 1 {\rm TeV}) &=& \sigma_{ZZ}(m_H=1 \; {\rm
TeV})
- \sigma_{ZZ}(m_H=0)  \simeq 14\; {\rm fb}  \nonumber \\
\sigma^{}_{\rm SEWS}({\rm LET}) &=& \sigma_{ZZ}(m_H=  {\infty})
- \sigma_{ZZ}(m_H=0)  \simeq7 \; {\rm fb} \nonumber
\end{eqnarray}
The signal rates are enhanced by about a factor $\sim$ 2--2.5
by increasing the CM energy from 1.5 to 2 TeV (compared with the
first numerical column in Table~\ref{tablei}).

There are other related works \cite{najima,kurihara} on the similar
subject to that discussed here. The results reported
here \cite{epemww} are updated and more optimistic.

\section{Heavy Higgs Boson And Strong $WW$
\hfill\break Scattering Signal at Photon Linear Colliders}

One can envision running a high energy linear collider in a $\gamma \gamma $
collision mode\footnote{This section was originally  drafted by M. Berger.}
by Compton backscattering laser beams off the electron
beams \cite{gkst}.
The feasibility of the heavy Higgs and strong vector boson scattering
signals is of immediate interest, and a comparison of the potential of the
$\gamma \gamma $ collider to that of the parent $e^+e^-$ collider is needed
to assess the motivation for adding the $\gamma \gamma $ option.
The central issue in all attempts to
find these signals at $\gamma \gamma $ colliders
is the size of the transversely polarized backgrounds
(which typically are much larger than the longitudinally polarized signals).

%The signals for the heavy Higgs
%($M_H\mathrel{\raisebox{-.6ex}{$\stackrel{\textstyle>}{\sim}$}}2M_Z$)
%and strong vector boson scattering
%have been extensively studied at hadron and $e^+e^-$ colliders.
%The heavy Higgs boson is of interest in its own right and one as
%heavy as 1 TeV has often been used as one model of the strong scattering.
The heavy Higgs boson (and other strong scattering signals)
at photon linear colliders can be investigated in the
modes \cite{Gunion,ggvv} $\gamma \gamma \to ZZ$, $\gamma \gamma \to W^+W^-$,
and \cite{brodsky}
$\gamma \gamma \to W^+W^-W^+W^-,W^+W^-ZZ$. The irreducible
backgrounds in the first two cases are severe, with
$\sigma(\gamma \gamma \to WW)\sim 90$ pb and
$\sigma(\gamma \gamma \to ZZ)\sim 250$ fb,
with the dominant contributions to the cross sections coming when both final
state bosons are transversely polarized. These contributions are strongly
peaked in the forward and backward regions, but are still very large even
after a cut on the scattering angle.
The full Standard Model calculations for the processes
$\gamma \gamma \to W^+W^-W^+W^-,W^+W^-ZZ$ have recently become available,
and the true size of the background can now be better ascertained.

The heavy Higgs boson in the channel $\gamma \gamma \to H \to ZZ$ is
analogous to
$gg \to H \to ZZ$ at hadron colliders, but the $W$-boson loops
give a much more substantial contribution at high energies
to the continuum background of the
former process than do the fermion loops that contribute to both processes.
The process \cite{jikia2} $\gamma \gamma \to ZZ$ is analogous to
``light-on-light''
scattering $\gamma \gamma \to \gamma \gamma $, but also includes the richer
structure associated with the longitudinal $Z$ bosons. The processes
$\gamma \gamma \to \gamma \gamma $ and $Z\to 3\gamma $ were previously
investigated, but in the kinematic regions where the $W$ boson loop
contributions are small.
These processes are very large at large center-of-mass
energies in photon-photon collisions \cite{jikia2,jikia3}.
These contributions arise
strictly from the nonabelian electroweak gauge theory and exist regardless
of whether a light Higgs boson exists or how the electroweak symmetry is
broken (in the
high energy limit the cross sections for the three
processes differ only by factors of $\tan \theta _W$ from the ratio of the
couplings of the $Z$ and the photon to the $W$).
At lower center-of-mass energies the cross section is more modest, and
the Higgs boson can be found at the $5\sigma $ level with 10 fb$^{-1}$ if
$M_H\mathrel{\raisebox{-.6ex}{$\stackrel{\textstyle<}{\sim}$}} 350$
GeV \cite{jikia2}. For higher masses the size of the Higgs boson peak
is reduced,
and the size of the continuum background increases rapidly. Therefore,
increasing the luminosity does not increase the reach in Higgs boson mass
significantly and not much is gained in the
$\gamma \gamma $ option over the $e^+e^-$ one, as far as discovery of the
Higgs boson in this channel
\footnote{However the $H\gamma \gamma $ coupling
can be measured for
$M_H\mathrel{\raisebox{-.6ex}{$\stackrel{\textstyle<}{\sim}$}} 350$ GeV
and could possibly shed light on the properties of the Higgs boson(s) and
heavy quanta that contribute to this coupling \cite{lighth}.}.

In both $\gamma \gamma \to ZZ$ and $\gamma \gamma \to W^+W^-$
the $s$-channel Higgs boson signal and the continuum background interfere
constructively for center-of-mass energies below the Higgs boson pole, and
interfere destructively above it.
Morris \etal \cite{mtz}\ have considered the possibility that the
Higgs boson signal
in $\gamma \gamma \to H \to W^+W^-$ can be seen as a resonance dip in the
$M_{WW}$ spectrum. Assuming a rather optimistic experimental resolution
$\sigma _{\rm res}=5$ GeV for $M_{WW}$,
they find a statistical significance of the
Higgs boson signal in excess of $5\sigma $ only for
$M_H\mathrel{\raisebox{-.6ex}{$\stackrel{\textstyle<}{\sim}$}} 175$ GeV at a
$\sqrt{s_{e^+e^-}}^{}=500$ GeV collider with 20 fb$^{-1}$ luminosity.
Assuming instead a luminosity of 100 fb$^{-1}$ one might expect
a $5\sigma $ effect for
$M_H\mathrel{\raisebox{-.6ex}{$\stackrel{\textstyle<}{\sim}$}} 300$ GeV.
With decreased experimental resolution of $M_{WW}$, the resonance dip will
become less statistically significant;
in any case the process $\gamma \gamma \to H \to ZZ$
discussed above has a higher reach.

In models with strong electroweak symmetry breaking the process
$\gamma \gamma \to Z_LZ_L$ can proceed more generally
through $\gamma \gamma \to W^+_LW^-_L$
followed by the rescattering $W^+_LW^-_L\to Z_LZ_L$. This is analogous to
the ordinary hadronic process $\gamma \gamma \to \pi^0 \pi^0$.
Early enthusiasm for the possibility of observing strong $WW$ scattering in
this way at a photon linear collider has diminished somewhat with the
observation by
Jikia that the large Standard Model background of transversely polarized
$Z$ pairs is overwhelming, just as $\gamma \gamma \to W^+_LW^-_L$ is much
smaller than $\gamma \gamma \to W^+_TW^-_T$.
Nevertheless, one can attempt to employ techniques to isolate the
longitudinal signal.
Since longitudinal and transverse $Z$ bosons can never be separated on an
event-by-event basis, one must resort to the application of statistical means
to try to separate the signal from the background. One such attempt made
recently \cite{bc} is to make use of the harder $p_T$ spectrum of the
longitudinal $Z$ decay products,
which has proved effective at hadron colliders. Unfortunately this
technique appears insufficient to provide a viable signal.
The strong scattering models employed smoothly extrapolated
the threshold behavior described by the low energy theorems in a manner that
satisfies unitarity. The result is that the transverse background is so
large that the nonresonant $J_Z=0$ contributions from the low energy theorem
are never observable, even with 100 fb$^{-1}$. Furthermore the
signal-to-background ($S/B$) ratio remains at only a few percent, so that
even with much higher luminosity there remains a potential problem with
systematic errors.

If one considers energies much in excess of one TeV, then one does not expect
the $s$-wave contribution to the scattering to necessarily dominate. The
mode $\gamma \gamma \to ZZ$ (and $\gamma \gamma \to W^+W^-$) couples to
$J_Z=0$ and $J_Z=2$, and resonances might ultimately be observable in these
channels. In fact
the QCD data has a prominent tensor resonance,
$\gamma \gamma \to f_2(1270) \to \pi^0\pi^0$. As a simple model one can
scale up the QCD data
by a factor of $v/f_{\pi}\sim 2600$, one sees that an analogous resonance
would appear in $\gamma \gamma \to Z_LZ_L$ at 3.4 TeV.
This tensor could be seen at a photon-photon collider
with 100 fb$^{-1}$ provided the energy of the collider is sufficient
($\sqrt{s_{e^+e^-}}=4$ TeV).
If a photon-photon collider can be made sufficiently monochromatic, the
luminosity requirement is not as severe (since this is a resonance).

It was proposed \cite{brodsky} that the strong $WW$ dynamics would
be better observed at a photon linear collider in the processes
$\gamma \gamma \to WWWW, ZZWW$, analogous to the strong scattering
process \cite{changail} $qq \to qqWW$.  One is still confronted with
the possibility that the transverse background is too large; this
issue has been addressed recently by Jikia \cite{jikia} and
Cheung \cite{cheung}.  Jikia finds that with luminosity of 200
fb$^{-1}$, a 1.5 TeV $e^+e^-$ collider operating in the $\gamma \gamma$
mode can successfully observe a heavy Higgs boson with mass up
to 700 GeV.  To obtain a reach in the Higgs boson mass up to 1 TeV,
a 2 TeV $e^+e^-$ collider is required. These results were obtained
taking into account the usual photon luminosity spectrum \cite{gkst}.
Jikia also finds that a 2 TeV linear collider in the $\gamma \gamma
$ mode is roughly equivalent to a 1.5 TeV $e^+e^-$ collider.
Cheung's conclusions are more optimistic, primarily because he
assumes a monochromatic photon spectrum. He finds that strong
scattering (including a 1 TeV Higgs boson) can be seen with a
$\gamma \gamma$ collider at a 2 TeV $e^+e^-$ collider with 100
fb$^{-1}$, and that at a 2.5 TeV $e^+e^-$ collider one would need
only 10 fb$^{-1}$ when the photon-photon spectrum is monochromatic
with $\sqrt{s_{\gamma \gamma}}\sim 0.8\sqrt{s_{e^+e^-}}$.  Since the
cross section of the signal rises rapidly with energy, one loses
statistical significance when it is convoluted with the usual photon
luminosity spectra obtained assuming zero conversion distance.  The
viability of this signal still requires that the decays of the $W$
and $Z$ bosons be incorporated, and detector simulations be
performed to determine realistic acceptances.

There is the possibility that the achievable luminosity of the
$\gamma \gamma $ collider may in practice
be much larger than that for a $e^+e^-$ collider.
If the necessary luminosities at the TeV $e^+e^-$ collider are
not attainable because of beam-beam interactions,
the $\gamma \gamma $ option might be the only way to achieve the needed
event rates.

\section{Summary}

With our present lack of experimental information on
the electroweak symmetry breaking mechanism, a
strongly-interacting electroweak sector remains a logical possibility.
In designing experiments at future high energy colliders, it is
necessary to carry out comprehensive studies for physics of the
strongly-interacting ESB sector, which has unique characteristics
in terms of experimental searches and special demands on the
detector performance.

In this report, we presented phenomenological discussions for
a strongly-interacting ESB sector at future $e^+e^-$ (NLC),
$p\bar p, pp$ (the upgraded Tevatron and LHC),
and $\gamma \gamma$ colliders. We adopted a (relatively) model
independent approach based on effective (chiral) Lagrangians.
We studied the cases in which the underlying dynamics is dominated
by a spin-zero, isospin-zero resonance; by a spin-one,
isospin-one resonance; and when there is no resonance at all
below 1~to~1.5~TeV.

%DiTeV:
$\bullet$ If new, strongly-interacting (techni-$\rho$-like),
vector multiplets exist, a 4 TeV upgraded Tevatron could be capable of
finding them for masses up to $M_V\sim 1$ TeV.
% \cite{besstev}.

%LHC all channel:

$\bullet$
The LHC has statistically significant sensitivity
to explore strongly interacting ESB physics.
For the gold-plated, purely leptonic decays
of the final-state $W$'s, and imposing stringent leptonic cuts,
forward-jet-tagging and central-jet-vetoing, large
SM backgrounds can be effectively suppressed. This has
been supported by realistic detector simulations.
Complementarity has been explored for the $W^\pm Z$ and $W^\pm W^\pm$
channels in studying a vector-dominance model or a non-resonant
model. A systematic comparison of the different final states
allows one to distinguish between different models to a
certain degree.
A statistically significant signal can be obtained for every
model (scalar, vector, or non-resonant)
in at least one channel with a few (1-3) years of running
(at an annual luminosity of 100 fb$^{-1}$) if the systematics
in the high luminosity environment are under control.
It is also demonstrated with detector simulations that
the semileptonic decays of a heavy Higgs boson,
$H \to W^+W^- \to l \nu jj$ and $H \to ZZ \to l^+ l^- jj$,
can provide statistically significant signals for $m_H=1$ TeV,
after several years of running at the high luminosity.

% Barklow's:
$\bullet$
The process $e^+e^- \to\ww$ is an effective probe of
strong electroweak symmetry breaking, especially for
physics with a vector resonance. With $\sqrt {s_{ee}}$=1.5 TeV
and 190 fb$^{-1}$, it should be possible to distinguish
the effects from a very heavy techni-$\rho$,
as well as a non-resonant amplitude, and from the standard model
with a light Higgs boson.
%
%There is also some sensitivity to
%the underlying dynamics in the case of a non-resonant
%amplitude ($L_{9L}$ and $L_{9R}$ in the chiral Lagrangian).

% Han's e+e- to WW fusion:
$\bullet$
The $WW$ fusion processes are complementary to the $s$-channel
$W^+W^-$ mode since they involve more  spin-isospin channels
(the $I=0$ and $I=2$ channels).
For an $e^+e^-$ collider at $\sqrt s=1.5$~TeV and 190 fb$^{-1}$,
the $W^+W^-/ZZ$ event ratio can be a sensitive probe of
a strongly-interacting electroweak sector.
%Indeed, the differences between the various models are quite
%marked.
% and the observation of such signals would provide strong indications
%about the underlying dynamics of the strongly-interacting EWSB.
%In fact, not only the ratio but also the size of the separate $W^+W^-$ and
%$ZZ$ signals contains valuable dynamical information.
%These results show
Statistically significant signals are found for a 1 TeV scalar or
vector particle.  There is also about a 6$\sigma$ signal for non-resonant LET
amplitudes via the $W^+W^- \to ZZ$ channel alone.
% without  the improvement by beam polarization.
For an $e^-e^-$ collider with the same energy and luminosity,
the LET signal rate  for the $\nu\nu W^-W^-$ ($I=2$) channel
is similar to the LET result of  $e^+e^-\to\bar\nu\nu ZZ$, as anticipated,
while the background rate is higher.
The signals are doubled for an $e^-_L$ polarized beam (or
quadrupled for two $e^-_L$ beams), whereas the backgrounds increase
by smaller factors.
%Hence polarization improves the significance
%of signals substantially, for a given luminosity.
%
%The signals also increase strongly with the CM energy.
A 2 TeV $e^+e^-$ linear collider would increase the signal rates
by roughly a factor of 2--2.5.

%gamma gamma:
$\bullet$
The heavy Higgs boson (and other strong scattering signals)
can be investigated at photon linear colliders in the
modes $\gamma \gamma \to ZZ$, $\gamma \gamma \to W^+W^-$,
and $\gamma \gamma \to W^+W^-W^+W^-,W^+W^-ZZ$.
The irreducible backgrounds in the first two cases are severe, while
the processes $\gamma \gamma \rightarrow WWWW, ZZWW$ seem to be more
promising. It is found that
to obtain a reach in the Higgs boson mass
up to 1 TeV, a 2 TeV $e^+e^-$ collider
is required with luminosity of 200 fb$^{-1}$. The results were obtained
taking into account the back-scattered photon luminosity spectrum.
It is also found  that such a 2 TeV
linear collider running in the $\gamma \gamma $ mode is roughly
equivalent to a 1.5 TeV $e^+e^-$ collider for the purpose of studying
strongly-interacting ESB physics (the correspondence scales roughly as
$\sqrt{s_{\gamma \gamma}}\sim 0.8\sqrt{s_{ee}}$).

% Golden's hidden:
$\bullet$
It is possible for the $W$ and $Z$ scattering
amplitudes to be small and structureless. If the symmetry breaking sector
contains a large number of particles in addition to the longitudinal gauge
bosons, there may be light but very broad ``resonances''.  A toy model
illustrating this scenario may be constructed using scalar $O(N)$ field theory
solved in the limit of large $N$.  This model contains a Higgs boson which,
due to its anomalous width, may be ``hidden'': the theory may be strongly
coupled without the number of two gauge boson events being large. In such
a scenario, the broad Higgs resonances may be difficult to observe
through the dominant hadronic final state at a hadron collider, but may
be possible at an $e^+e^-$ linear collider because of a cleaner
experimental environment.

\section{Acknowledgment}

We are grateful to many contributors
who have provided materials to this report.
T. Barklow provided a figure of Ref.~\cite{barklow};
M. Berger drafted the original version of Section 5
on the photon linear collider;
R. Casalbuoni, P. Chiappetta, A. Deandrea, S. De Curtis,
D. Dominici, R. Gatto contributed figures and discussions
of Ref.~\cite{besstev,lin} for the BESS model results;
K. Cheung and G. Ladinsky offered assistance regarding
Ref.~\cite{baggeretali};
M. Herrero provided a large list of references;
and Y. Kurihara provided informative discussions on Ref.~\cite{kurihara}.
We also express our gratitude to many of the authors of the papers
cited in the
references, in particular to our collaborators, for helpful discussions
and comments and for helping us collect all the relevant literature for
this review. Finally, we would like to thank M. Drees
for carefully reading an early draft.

\vfil\eject

\end{document}